\documentclass[aps,preprint]{revtex4}
\usepackage{color,graphicx}
\usepackage[normalem]{ulem}
\usepackage{bm}
\usepackage{times}
\usepackage{graphics}
\usepackage{dcolumn}
\usepackage{bm}
\usepackage{xcolor}
\usepackage{dcolumn}
\usepackage{bm}
\usepackage{times}
\usepackage{amsmath}
\usepackage{amssymb}
\makeatletter
\newenvironment{figurehere}
{\def\@captype{figure}}
{}
\makeatletter

\begin{document}
\newcommand{\red}[1]{\textcolor{red}{#1}}
\newcommand{\green}[1]{\textcolor{green}{#1}}
\newcommand{\blue}[1]{\textcolor{blue}{#1}}
\newcommand{\YRS}{YbRh$_2$Si$_2$}
\newcommand{\TA}{$T_{\textrm{A}}$}
\newcommand{\TB}{$T_{\textrm{B}}$}
\newcommand{\TC}{$T_{\textrm{c}}$}
\newcommand{\TN}{$T_{\textrm{AF}}$}
\newcommand{\replace}[2]{\sout{#1} \textcolor{red}{#2}}
\newcommand{\rereplace}[2]{\sout{#1} \textcolor{blue}{#2}}
\newcommand{\frank}[2]{\sout{#1} \textcolor{red}{#2}}
\newcommand{\qs}[2]{\sout{#1} \textcolor{magenta}{#2}}
\newcommand{\TF}{$T_{\textrm{F}}$}
\newcommand{\BN}{$B_{\textrm{N}}$}
\newcommand{\BC}{$B_{\textrm{c}}$}
\newcommand{\BA}{$B_{\textrm{A}}$}
\newcommand{\BB}{$B_{\textrm{B}}$}
\newcommand{\xc}{$x_{\textrm{c}}$}
\newcommand{\TLFL}{$T_{\rm FL}$\,}
\newcommand{\TK}{$T_{\rm K}$\,}
\newcommand{\cp}{$c_{p}(T)$\,}
\newcommand{\cpT}{$c_{p}(T)/T$\,}
\newcommand{\vF}{$v_{\rm F}$\,}
\newcommand{\lel}{$l_{\rm el}$\,}
\newcommand{\et}{~\textit{et al.\,}}
\newcommand{\hlr}[1]{{\textcolor{red}{#1}}}
\newcommand{\LO}{L$_{0}$}
%
%
\begin{center}
{\large\bf Emergence of superconductivity in the canonical heavy-electron metal \YRS}\\
[0.5cm]
Erwin Schuberth$^{1,2\,*}$, Marc Tippmann$^{1}$, Lucia Steinke $^{1,2}$, Stefan Lausberg$^{2}$, 
Alexander Steppke${^2}$, Manuel Brando$^{2}$, Cornelius Krellner$^{2,3}$, 
Christoph Geibel$^{2}$, Rong Yu$^{4}$, Qimiao Si$^{5\,*}$, and Frank Steglich$^{2,6,7\,*}$\\
{\textit{$^{1}$Walther Meissner Institute for Low Temperature Research,
Bavarian Academy of Sciences, 85748 Garching, Germany}\\
\textit{$^{2}$Max Planck Institute for Chemical Physics of Solids,
01187~Dresden, Germany}\\
\textit{$^{3}$Physics Institute,  University of Frankfurt, 
60438 Frankfurt, Germany}\\
\textit{$^{4}$Department of Physics, Renmin University of China, Beijing 100872, China.}\\
\textit{$^{5}$Department of Physics and Astronomy, Rice University,
Houston, TX 77005, USA}\\
\textit{$^{6}$Center for Correlated Matter, Zhejiang University, Hangzhou, Zhejiang 310058, China}\\
\textit{$^{7}$Institute of Physics, Chinese Academy of Sciences, Beijing 100190, China}\\
}

\vspace{0.5cm}
$^{*}$To whom correspondence should be addressed.\\
E-mail: steglich@cpfs.mpg.de, eschuber@ph.tum.de, qmsi@rice.edu.
\end{center}
\vspace{0.8cm}

\textbf{One-sentence summary}:
We demonstrate that heavy-electron superconductivity develops in \YRS\, due to the weakening of its antiferromagnetism by the ordering of nuclear spins, providing evidence that quantum criticality is a robust mechanism for unconventional superconductivity.

\vspace{0.5cm}
\textbf{We report magnetic and calorimetric measurements down to $T = 1$\,mK on the canonical heavy-electron metal \YRS. The data reveal the development of nuclear antiferromagnetic order slightly above 2\,mK. The latter weakens the primary electronic antiferromagnetism, thereby paving the way for heavy-electron superconductivity below \TC\ = 2\,mK. Our results demonstrate that superconductivity driven by quantum criticality is a general phenomenon.}

Unconventional (i.e., non-phonon mediated) superconductivity, which has been attracting much interest since the early 1980s, is often observed at the border of antiferromagnetic (AF) order~\cite{cuprates}. As exemplified by heavy-electron (or heavy-fermion) metals, the suppression of the AF order opens up a wide parameter regime where the physics is controlled by an underlying quantum critical point (QCP)~\cite{Loehneysen07,Gegenwart}. A central question, then, concerns the interplay between quantum criticality and unconventional superconductivity in strongly correlated electron systems such as heavy-electron metals. In many of the latter superconductivity turns out to develop near such a QCP~\cite{Loehneysen07,Gegenwart,Mathur}. However, the absence of superconductivity in the prototypical quantum critical material \YRS\ (Ref.~\cite{Custers}) has raised the question as to whether the presence of an AF QCP necessarily gives rise to the occurrence of superconductivity. Because \YRS\ exists in the form of high-quality single crystals, it is meaningful to address this issue at very low temperatures without seriously encountering the limitations posed by disorder. We have therefore used this heavy-electron compound to carry out the first study on quantum critical metals at ultra-low temperatures.

\YRS\ exhibits AF order below a N\'{e}el temperature $T_{\textrm{AF}} = 70$\,mK. A small magnetic field of $B = 60$\,mT, when applied within the basal plane of the tetragonal structure, continuously suppresses the magnetic order and induces a QCP, presumably of unconventional nature~\cite{Paschen,Friedemann}. Electrical resistivity measurements down to 10\,mK have failed to show any indications for superconductivity~\cite{Custers}. Recognizing that a critical field of 60\,mT is unlikely to sustain even heavy-electron superconductivity with a \TC\, of less than 10\,mK, a different means of suppressing the antiferromagnetism is needed to eventually reveal any potential superconductivity at its border.

We take advantage of the early recognition that hyperfine coupling to nuclear spins can considerably influence the electronic spin properties near a quantum phase transition \cite{Ronnow}. Furthermore, measurements on PrCu$_2$ and related compounds have demonstrated a large coupling between the electronic and nuclear spins in rare-earth-based intermetallics at temperatures as high as 50 mK~\cite{Andres,Steinke}. These considerations raise the possibility of using the presence of nuclear spins to weaken the electronic AF order, thereby enabling the formation of a superconducting state. We note that the application of pressure is unsuccessful to reach a QCP in an AF Yb-based material, as this will strengthen the magnetic order - opposite to the case of Ce-based systems where magnetism usually becomes weakened by pressure. While Ce does not exhibit a nuclear spin, two of the Yb isotopes have finite nuclear-spin values, see below and Sec.~F of SOM.

We have carried out magnetic and calorimetric measurements on high-quality \YRS\, single crystals, using a nuclear-demagnetization cryostat with a base temperature of 400\,$\mu$K (see SOM). Figs.~\ref{fig1}A and~\ref{fig1}B display the temperature dependence of the field-cooled (fc) DC-magnetization $M(T)$, measured upon warming at various magnetic fields $B$ ranging from 0.09\,mT up to 25\,mT, applied within the basal plane of the \YRS\ single crystals. The curves display peaks at 70\,mK, which is the well-established N\'{e}el temperature for the AF order, as well as additional low-temperature anomalies. There is a second peak in $M(T)/B$ at \TC\ $\simeq 2$\,mK which indicates the almost simultaneous onset of a nuclear-dominated AF order (``A phase'') and the Meissner effect (see below). It is visible above 1\,mK up to 23\,mT and had already been observed before~\cite{Schuberth13}. In addition, there is a shoulder around \TB\, $\approx 10$\,mK. As shown in Fig.~\ref{fig1}C, the results of the fc and zero-field-cooled (zfc) measurements become different below \TB. This disparity, which is ascribed to superconducting fluctuations (see Sec.~F, SOM), can be followed as a function of the magnetic field, up to the limit of $B = 0.5$\,mT of our setup for measurements of the DC-magnetization cooled at zero field.

At $T \simeq 2$\,mK, the zfc DC-$M(T)/B$ (0.012\,mT) shows a sharp increase upon warming, starting from negative values (Fig.~\ref{fig1}C). This indicates a substantial shielding signal due to superconductivity. Raising the temperature further, the zfc $M(T)/B$ slowly increases until at 10 mK it meets the fc curve. To verify this finding, we have carried out measurements of the AC-susceptibility, $\chi_{\textrm{ac}}$, under nearly zero-field conditions (see Sec.~D, SOM). As shown in Fig.~\ref{fig1}D, its real part, $\chi'_{\textrm{ac}}(T)$, displays an even more pronounced diamagnetic signal, larger than what was found for the canonical heavy-electron superconductor CeCu$_{2}$Si$_{2}$ (Ref.~\cite{Steglich}), again confirming the occurrence of superconducting shielding. In addition, the reduction of the fc magnetization upon cooling to below 2\,mK reflects flux expulsion from the sample (Meissner effect). The relatively small Meissner volume of $\approx 3$\,\% is most likely due to strong flux pinning (see Sec.~C, SOM). As shown in Fig.~S6 the superconducting phase transition is of first order. This suggests that superconductivity does not coexist on a microscopic basis with AF order, as previously observed for A/S-type CeCu$_{2}$Si$_{2}$, cf. Sec.~D of SOM.

In Fig.~\ref{fig2}A, the specific heat is displayed as $C(T)/T$ at $B = 2.4$\,mT and 59.6\,mT, respectively. As the electronic specific heat can be completely neglected below $T \approx 10$\,mK (Ref.~\onlinecite{Steppke}), $C(T)$ denotes the nuclear contribution in this low-$T$ regime. In addition, we show the calculated nuclear specific heats at various fields from Ref.~\cite{Steppke}, which include the quadrupolar as well as the Zeeman terms. At zero field the nuclear specific heat is completely dominated by the nuclear quadrupole states, to which the Zeeman terms due to the nuclear spin states add at $B > 0$. In Fig.~\ref{fig2}B, we display $\Delta C(T)/T$ where $\Delta C$ marks the difference between the specific heat measured at the lowest field $B = 2.4$\,mT and the nuclear quadrupole contribution calculated for $B = 0$ (Ref.~\cite{Steppke}). Our $\Delta C(T)/T$ results clearly reveal a peak at $T \approx 1.7$\,mK. Assuming a continuous phase transition, the transition temperature can be obtained by replacing the high-$T$ part of this peak by a sharp jump and at the same time keeping the entropy unchanged. This yields a jump height of about 1000\,J/K$^2$mol and \TA\ ($B = 2.4$\,mT) $\approx 2$\,mK, almost coinciding with the superconducting transition temperature \TC\, from the magnetic measurements discussed above (Fig.\ref{fig1}). Because the effect of the magnetic field on the quadrupole contribution to the nuclear specific heat is of higher order only, we can use the $\Delta C(T)/T$ data of Fig.~\ref{fig2}B to estimate the nuclear spin entropy (at $B = 2.4$\,mT), $S_{\rm {I}}(T)$, see Sec.~F, SOM. $S_{\rm {I,tot}} \simeq 1.35R\ln2$, the total nuclear spin entropy of \YRS\ for $B$ = 2.4\,mT, is reached at $T \approx$ 10\,mK where  $\Delta C(T)$ vanishes within the experimental uncertainty (see Fig.~\ref{fig2}C). Upon cooling to $T$ = \TA, $S_{\rm {I}}(T)$ decreases to about $0.94 S_{\rm {I,tot}}$; i.e., most of this nuclear spin entropy must be released below the phase transition temperature \TA. While the entropy due to the $^{103}$Rh and $^{29}$Si spins is temperature independent at $T > 1$\,mK, the Yb-derived spin entropy $S_{\rm Yb}(T)$ decreases by 26\,\% upon cooling from 10 to 2\,mK (cf. Sec. F, SOM). This indicates substantial short-range order, consistent with a second-order (antiferro)magnetic phase transition. We stress that this huge entropy at ultra-low temperatures can only be understood if the ordering transition at $T_{A}$ involves the Yb-derived nuclear spins to a large degree.

As shown below, the A phase forming at \TA\ $\simeq 2$\,mK is an electronic-nuclear hybrid phase which is dominated by the Yb-derived nuclear spin ordering. But, it also contains a small (1-2\,\%) $4f$-electronic component, which contributes about 1/3 to the decrease in $M(T)$ below \TA. Since the nuclear phase transition cannot be resolved because of the very small nuclear moment, the major part of this reduction of $M(T)$ ($\sim 2/3$) must be due to the Meissner effect (Sec.\,C, SOM). A measurement of the fc DC-magnetization at very low fields reveals two separated phase transitions close to $T = 2$\,mK: \TA\, $>$ \TC\, (see Fig.~S3B, SOM). Upon increasing the field up to about 3 -- 4\,mT, however, they appear to merge within the experimental uncertainty. This peak in the fc DC-$M(T)$ curve remains visible (above 1\,mK) up to $B \simeq 23$\,mT. By analyzing magnetization data taken between 0.8 and 540\,mK at a field of 10.1\,mT we conclude that superconductivity is likely to exist and concur with the A-phase at elevated fields as well, consistent with the evolution of the $M(T)$ peak as a function of field (Sec. C, SOM).

The huge initial slope of the superconducting upper critical field $B_{\textrm{c2}}(T)$ at $T_{\textrm{c}}$ $\simeq 25$\,T/K from both shielding (inset of Fig.~\ref{fig3}) and Meissner measurements (Fig. S3C, SOM) corresponds to an effective charge-carrier mass of several 100\,$m_{\textrm{e}}$ ($m_{\textrm{e}}$ being the rest mass of the electron), implying that the superconducting state is associated with the Yb-derived $4f$ electrons (``heavy-electron'' superconductivity). Extrapolating the positions of the low-temperature  fc $M(T)$ peak to zero temperature, the critical field of the A phase $B_{\textrm{A}} = B(T_{\textrm{A}} \rightarrow 0)$ is found to be 30 -- 60\,mT which corresponds to an effective electronic $g$-factor $g_{\textrm{eff}} = k_{\textrm{B}} T_{\textrm{A}}(B = 0) / \mu_{\textrm{B}} B_{\textrm{A}} =$ 0.03 -- 0.06, much smaller than the in-plane electronic $g$-factor, 3.5 (Ref.~\cite{Sichelschmidt}), but a factor of 20 to 40 larger than in case of a  purely nuclear-spin ordering transition. We can understand this $g_{\textrm{eff}}$ if the ordered moment is a hybrid of the electronic and nuclear spins with, at most, 2\,\%  of the ordered moments being associated with the $4f$-electron derived spins.

In order to explore the role of the nuclear spins in the phase diagram, we have carried out a Landau theory of the interplay between the magnetic orders of the electronic and nuclear spins. Consider the electronic AF order, with an order-parameter $m_{\rm AF}$ at the AF wavevector ${\boldsymbol {Q}}_{\rm AF}$, as well as two bilinearly coupled order parameters, $m_{\rm J}$  and  $m_{\rm I}$, the staggered magnetizations of the electronic and nuclear spins at another finite wavevector ${\boldsymbol Q}_1 \ne {\boldsymbol Q}_{\rm AF}$. The bilinear coupling arises from the hyperfine coupling between the two order parameters having the same wavevector. The Landau theory will then have the following free energy functional:
\begin{eqnarray}
\label{eq1}
f &=& \frac{1}{2} r_{\rm AF} \phi^2_{\rm AF} + \frac{1}{4} u_{\rm AF} \phi^4_{\rm AF} + 
\frac{1}{2} r_{\rm J} \phi^2_{\rm J} + \frac{1}{4} u_{\rm J} \phi^4_J + \frac{1}{2} r_{\rm I} \phi^2_{\rm I} + 
\frac{1}{4} u_{\rm I} \phi^4_{\rm I} \nonumber\\
&& - \lambda \phi_{\rm J} \phi_{\rm I} + \frac{1}{2}\epsilon \phi^2_{\rm AF} \phi^2_{\rm I} + 
\frac{1}{2}\eta \phi^2_{\rm J} \phi^2_{\rm AF},
\end{eqnarray}
where $\phi_{\rm AF}$, $\phi_{\rm J}$ and $\phi_{\rm I}$ are respectively the normalized order parameters $m_{\rm AF}$, $m_{\rm J}$  and  $m_{\rm I}$, the $r$'s are quadratic couplings, and the $u$'s as well as $\epsilon$ and $\eta$ are the intra-component as well as inter-component quartic couplings (see Sec.~G, SOM).

Under suitable conditions (see SOM), this leads to two stages of phase transitions, as shown in Fig.~\ref{fig4}. The phase transition at $T_{\textrm {AF}}$ corresponds to the primary AF order setting in at about $70$ mK, and is not much affected by the nuclear spins. In a suitable parameter range of the Landau theory, the nuclear $\phi_{\rm I}$ order dominates over the electronic $\phi_{\rm J}$ order and, furthermore, suppresses the primary electronic $\phi_{\rm AF}$ order. A second transition occurs at $T_{\rm hyb}$, which represents a hybrid electronic-nuclear spin order. The component that is associated with the nuclear spins generates substantial entropy for the transition, which provides the understanding of the large nuclear spin entropy that is experimentally observed (see Fig.~\ref{fig2}C and Sec.~E, SOM). In addition, the effective $g$-factor is approximate to $g_{\rm el} \phi_{\rm J}/\phi_{\rm I}$, which is substantially smaller than the bare $g$-factors for the $4f$-­electrons. This allows us to understand the $g_{\rm eff} < 0.1$ observed in the experiment.  

Fig.~\ref{fig4} describes the two stages of transitions. Below $T_{\rm AF}$, the N\'{e}el order develops. The growth of the N\'{e}el order parameter, $m_{\rm AF}$, is arrested as the temperature is lowered through $T_{\rm hyb}$, due to the onset of the nuclear spin order. This diminished $m_{\rm AF}$ places the electronic phase to the regime close to the QCP that underlies the pure electronic system in the absence of any hyperfine coupling. This quantum criticality effectively induced by the nuclear-spin order at zero magnetic field would naturally lead to the development of a superconducting state (see Sec.~I, SOM). As inferred from the experimental results, fluctuations of the A-phase set in already near \TB\ and lead to a substantial reduction of the staggered magnetization and the emergence of superconducting fluctuations well above the A-phase ordering temperature, see also the discussion in Sec. I, SOM.

The huge entropy near \TA\ $\gtrsim 2$\,mK is one of the most pronounced features in our observation. In addition to the ordering of the nuclear spins, which competes with the primary electronic order and thus paves the way for superconductivity, an intriguing alternative possibility for this entropy is the involvement of a ``nuclear Kondo effect'', i.e., the formation of a singlet state between the nuclear and conduction-electron spins. The resulting ``superheavy'' fermions may be assumed to form Cooper pairs and cause a superconducting transition at \TC\ $\approx 2$\,mK that would be probed by the magnetic as well as the specific-heat measurements. While our estimate of the nuclear Kondo temperature and the quasiparticle effective mass reveal discrepancies with this picture (see Sec. E, SOM), future theoretical and experimental work is highly welcome to investigate the possible role of the nuclear Kondo effect in generating superconductivity in \YRS.

As discussed in Sec.~H of the SOM, it is likely that the coupling of electronic and nuclear spin orders as well as the concomitant emergence of new physics is not a unique property of \YRS. Systematic studies of other heavy-electron antiferromagnets at ultra-low temperatures are required, to find out whether a hybrid electronic-nuclear order is potentially a general phenomenon. In addition, a comparative study would be highly welcome to check whether SC is indeed absent in isotopically enriched \YRS\,, without Yb-derived nuclear spins, similar to the one studied in Ref.~\cite{Knebel06}.

Superconductivity in heavy-electron metals is often discussed in terms of an effective electron-electron attractive interaction provided by nearly quantum critical fluctuations associated with a spin-density wave (SDW) QCP~\cite{Mathur,Monthoux}. This was recently exemplified, via inelastic neutron scattering, for CeCu$_2$Si$_2$ (Refs.~\cite{Arndt,Stockert}). On the other hand, in the special case of CeRhIn$_5$, superconductivity appears to form~\cite{Shishido,Park,Knebel08} in the vicinity of a Kondo-breakdown QCP~\cite{Si01,Coleman,Senthil}. It is likely that the same applies to $\beta$-YbAlB$_4$ (Ref.~\cite{Nakatsuji}). This is in contrast to the behavior of CeCu$_{6-x}$Au$_x$, the prototype heavy-electron metal which exhibits such a Kondo breakdown QCP~\cite{Loehneysen,Schroeder} but shows no superconductivity down to $T \approx 20$\,mK (Ref.~\cite{Loehneysen95}). In this case, it is natural to assume that unconventional superconductivity is, at least above 20 mK, suppressed by the alloying-induced 
disorder. By contrast, in high-quality single crystals of the antiferromagnet \YRS, another well-established heavy-electron metal with a Kondo breakdown QCP~\cite{Paschen,Friedemann}, our work shows that superconductivity does develop at \TC\ $= 2$\,mK. Here, the primary electronic order which appears to be detrimental to superconductivity is sufficiently weakened by the ordering of the nuclear spins; this in turn pushes the system close to the underlying QCP. The concomitant quantum critical fluctuations, rather than the magnon fluctuations as in the case of UPd$_{2}$Al$_{3}$ (Ref.~\cite{Sato}), are therefore the driving force for superconductivity. This heavy-electron superconductivity may be called "high \TC", in the sense that it is limited by an exceedingly high ordering temperature of nuclear spins. Moreover, the emergence of superconductivity in \YRS\ provides evidence for the notion that has been implicated by de-Haas-van-Alphen studies on CeRhIn$_5$ in high pulsed magnetic fields~\cite{Yuan}; namely, superconductivity is robust in the vicinity of such a Kondo-breakdown QCP, which may be considered a zero-temperature $4f$-orbital selective Mott transition. Therefore, these results provide a new link between the unconventional superconductivity of heavy-electron materials and that occurring near true Mott transitions, e.g., in the cuprates~\cite{cuprates} and organic charge-transfer salts~\cite{Kanoda}. Finally, our conclusion that quantum criticality is a robust mechanism for superconductivity pertains to wider settings such as finite-density quark matter~\cite{Alford}.
%
%
\\
\\
\vspace*{0.8cm}\noindent{\bf \large Acknowledgments} \newline
We are indebted to K. Andres, P. Coleman, P. Gegenwart, S. Paschen and S. Wirth for useful discussions and R. Gross for his support of the project at the WMI. Part of the work at Dresden was  supported by the DFG Research Unit 960 ``Quantum Phase 
Transitions''. Q.~S.\ was supported by NSF Grant DMR-1309531 and the Robert A.\ Welch Foundation Grant No.\ C-1411. E. S., Q. S. and F. S. are grateful to the hospitality of the Institute of Physics, Chinese Academy of Sciences, Beijing. Q. S. and F. S. acknowledge the support in part by the National Science Foundation under Grant No. 1066293 and the hospitality of the Aspen Center for Physics.
\newpage
\begin{figurehere}
\begin{center}
\includegraphics[width=0.9\columnwidth]{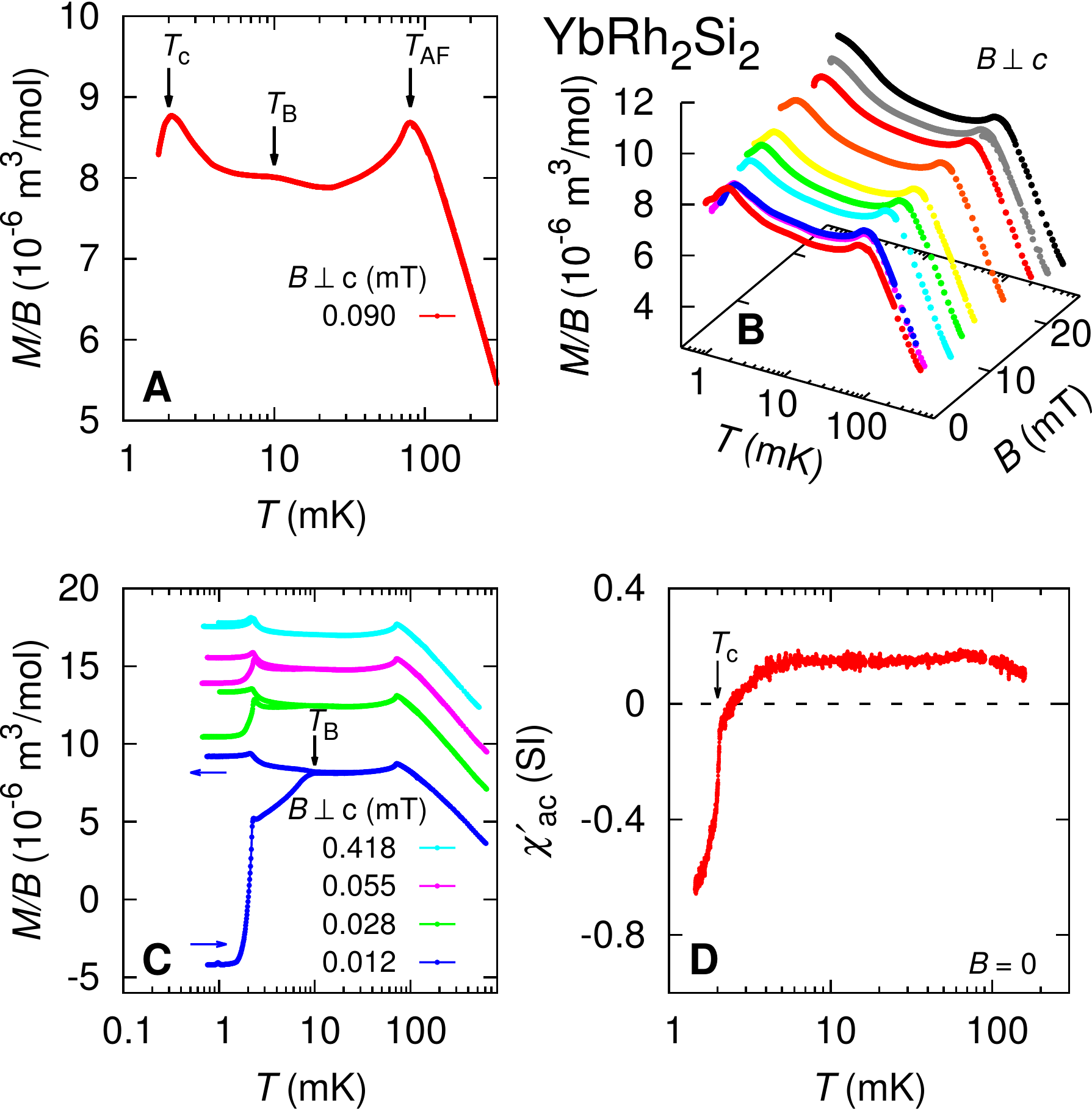}
\caption{\label{fig1}
Temperature dependence of the DC-magnetization at varying fields and AC-susceptibility at $B = 0$ for \YRS, see also SOM. (\textbf{A}) DC-magnetization curve of \YRS\ taken at $B = 0.09$\,mT applied within the basal plane. Three main features are clearly visible: the AF phase transition at \TN\ = 70\,mK, a shoulder in magnetization at \TB\ $\approx 10$\,mK and a sharp peak at \TC\ = 2\,mK. (\textbf{B}) Series of magnetization data taken at fields of 0.10, 1.13, 1.13, 5.01, 7.48, 10.12, 15.01, 20.04, 22.42 and 25.02\,mT. (\textbf{C}) Zero-field-cooled and field-cooled DC-magnetization traces taken at selected small magnetic fields. The traces at 0.028, 0.055 and 0.418\,mT were shifted upwards for better visibility. For the smallest magnetic field of 0.012\,mT a sharp diamagnetic shielding signal is observed, suggesting a superconducting phase transition. (\textbf{D}) The AC-susceptibility was measured using a SQUID magnetometer by modulating a primary coil around the pickup coils. Here we show the in-phase signal $\chi'_{\textrm{ac}}(T)$ (at 17\,Hz), after having compensated the earth field. All features seen in the DC-magnetization are detected by the AC-susceptibility at \TN, \TB\, and \TC, too. Importantly, the large negative values of the zero-field-cooled DC-magnetization at $B = 0.012$\,mT (Fig.~\ref{fig1}C) and of $\chi'_{\textrm{ac}}(T)$ indicate superconducting shielding, while the low-temperature peak in the field-cooled DC-magnetization (Figs.~\ref{fig1}A-C) signal the onset of the Meissner effect. Measurements in panels 1A, 1B, 1C and 1D were performed on samples \#1, \#2, \#3 and \#4, respectively.}
\end{center}
\end{figurehere}
\newpage
\begin{figurehere}
\begin{center}
\includegraphics[width=0.9\columnwidth]{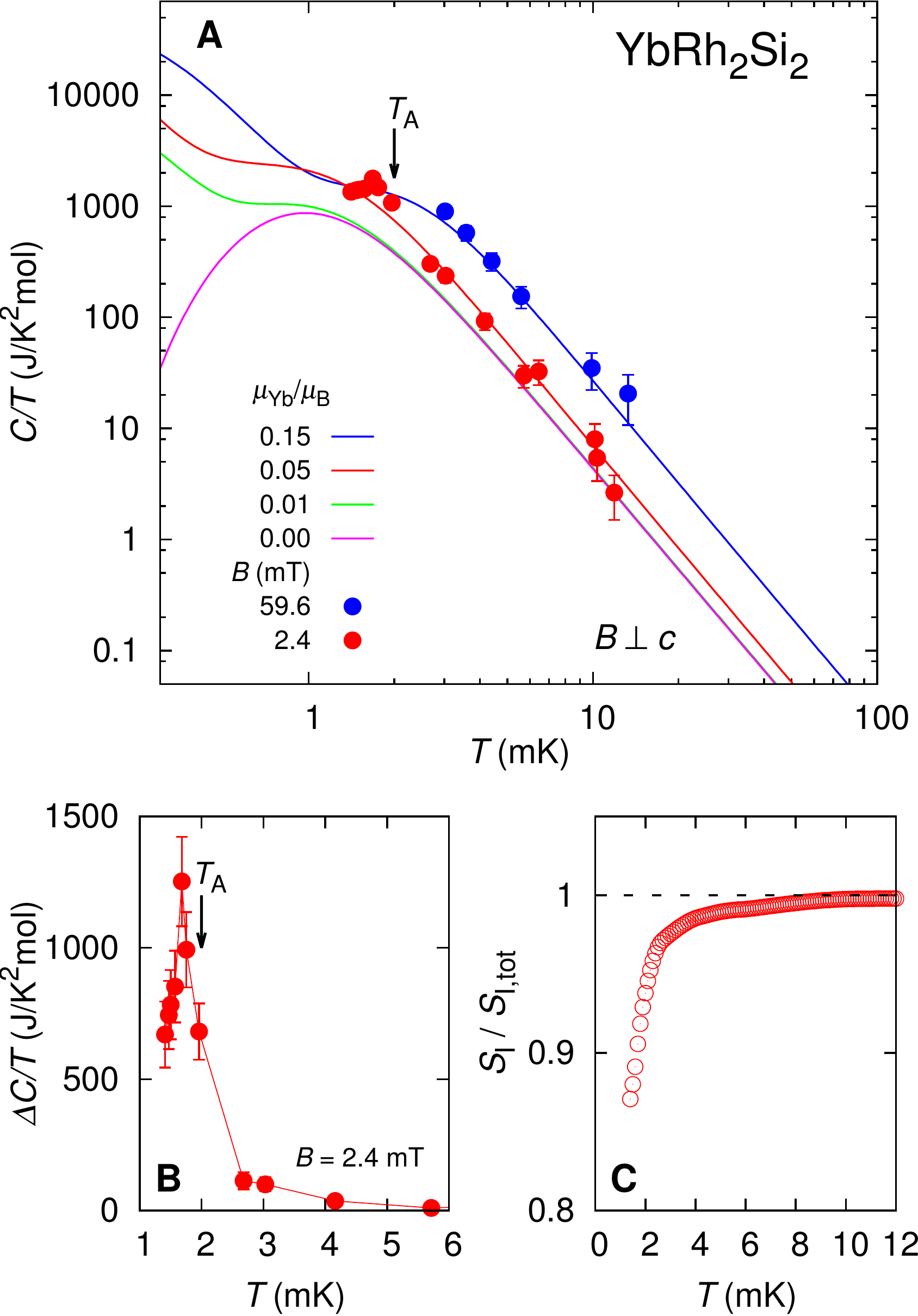}
\caption{\label{fig2}
Nuclear specific heat and entropy of \YRS. (\textbf{A}) The temperature dependence of the specific heat $C(T)$ of \YRS\ divided by $T$ is shown for $B$ = 2.4 and 59.6\,mT. $C(T)$ was measured with the semiadiabatic heat-pulse method using the sample itself as a thermometer along the DC-magnetization curve. The sample-to-bath relaxation time constants were short enough to allow us to use this method. Analyzing the recorded cooling curve, the specific heat at the same temperature was also determined from the relaxation time and the thermal resistance of the weak link (a thin high-purity Ag wire), the latter measured separately (see Sec.~E, SOM). The 2.4\,mT data extend down to 1.4\,mK. The solid lines denote the calculated nuclear specific heat from Ref.~\cite{Steppke}, which is the sum of the quadrupolar term and the Zeeman term with three selected field-induced Yb-derived ordered magnetic moments: 0.01, 0.05 and 0.15\,$\mu_{\textrm{Yb}}/\mu_{\textrm{B}}$. The large errors at temperatures above 10\,mK are due to the uncertainty in the subtraction of the addendum, see Sec.~E of SOM. (\textbf{B}) $\Delta C(T)/T$ was obtained by subtracting the nuclear quadrupolar contribution calculated for $B = 0$ from the data at $B = 2.4$\,mT. A peak in $\Delta C(T)/T$ occurs at $\approx 1.7$\,mK. Assuming the transition to be of second order, an equal-area construction yields a nuclear phase transition temperature \TA\ $\approx 2$\,mK. This coincides with the peak position found in the DC-magnetization (cf. Fig.~\ref{fig1}), i.e., the superconducting critical temperature \TC, at 2.4\,mT. The associated jump of $\Delta C(T)/T$ is of the order of 1000\,J/$\textrm{K}^{2}$ mol. The error bars reflect the statistical error in the measurements of the specific heat by utilizing a quasi-static heat-pulse technique as well as the relaxation method. In the latter case, the error bar contains the statistical error in determining both the relaxation time and the heat conductivity of the weak link. In total, for each field two runs have been performed and, therefore, four sets of data at the same temperature were used for determining the specific heat. Each data point was weighted by its reciprocal error. At the lowest temperatures, the error associated with the relaxation method is essentially smaller than that of the heat-pulse measurement. (\textbf{C}) From $\Delta C(T)/T$ (Fig.~\ref{fig2}B), a rough estimate can be made for the nuclear spin entropy $S_{\textrm{I}}(T)$ at $B = 2.4$\,mT, see text. We have normalized $S_{\textrm{I}}(T)$ to $S_{\textrm{I,tot}}$, the total nuclear spin entropy in \YRS\ at $B = 2.4$\,mT, reached at about 10 mK. Subtracting from $S_{\textrm I}(T)$ the contribution of the nuclear Si and Rh spins which is almost temperature independent at $T \geq$ 1mK, we obtain the corresponding values $S_{\textrm {Yb}}(T)$ for the nuclear Yb spins. It can be seen that a considerable portion (about 26\,\%) of $S_{\textrm{Yb,tot}}$ is released upon cooling to \TA\ $\simeq$ 2\,mK. Therefore, an entropy of $\approx 0.74 S_{\textrm{Yb,tot}}$ is expected to be released below \TA, i.e., inside the A phase (see Sec.~E, SOM). Measurements were performed on sample \#3.}
\end{center}
\end{figurehere}
\newpage
\begin{figurehere}
\begin{center}
\includegraphics[width=0.9\columnwidth]{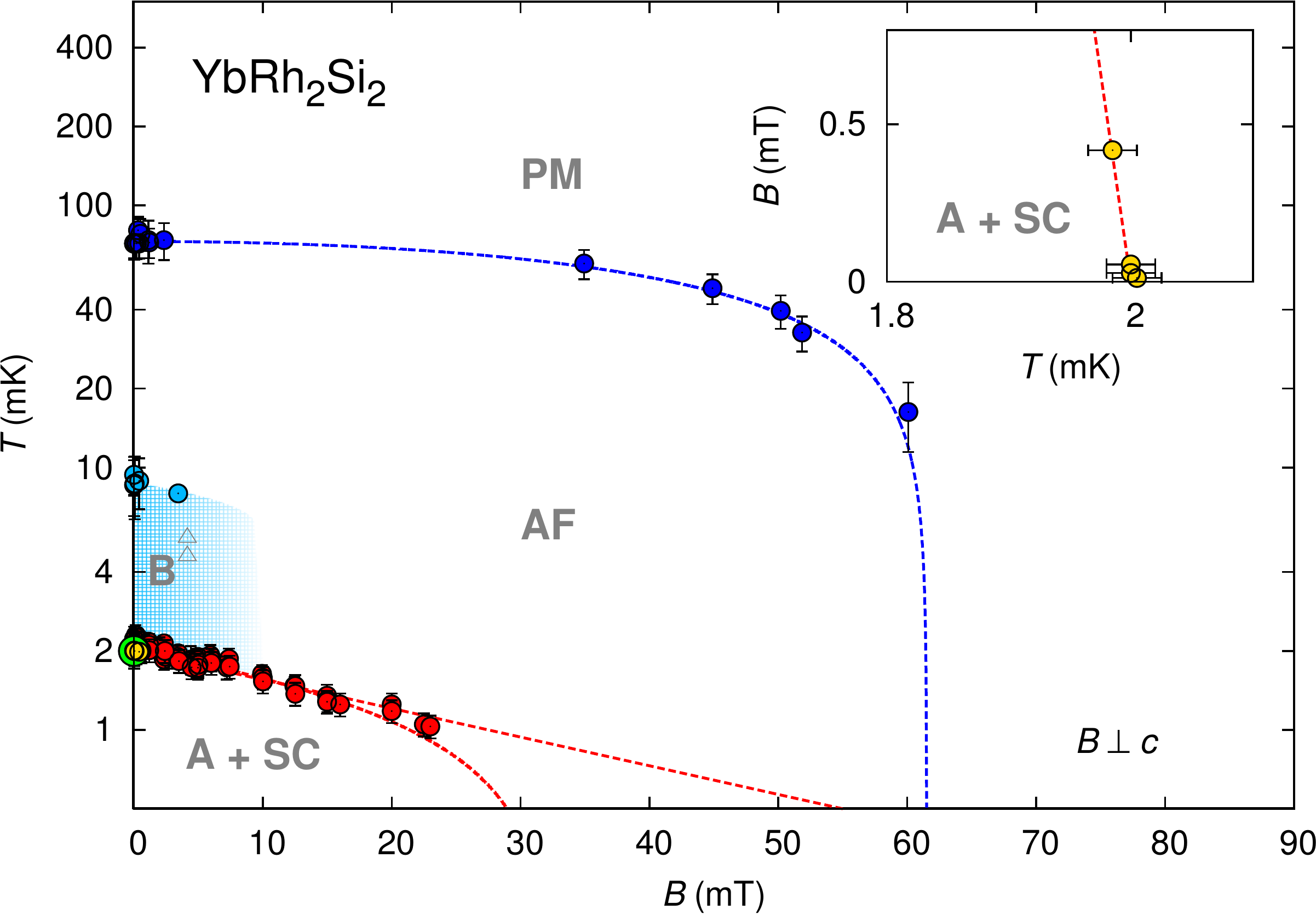}
\caption{\label{fig3}
Generic $T-B$ phase diagram of \YRS. This phase diagram is obtained from DC-magnetization and AC-susceptibility measurements in several magnetic fields. A total of 4 samples were measured and no significant sample dependence was found. AF indicates the electronic antiferromagnetic order (\TN\ = 70\,mK), PM indicates the paramagnetic state. All data points used to illustrate the AF -- PM phase boundary \TN$(B)$ were obtained in the present study; several more are available in previous work. In particular, it was shown earlier by magnetostriction measurements that the transition at \TN$(B)$ stays to be of second order to the lowest accessible temperature of 20\,mK (Ref.~\onlinecite{Gegenwart}). The hatched light-blu area indicates the onset of A-phase fluctuations which give rise to a reduction of the staggered magnetization and a splitting of the zero-field-cooled and the field-cooled DC-magnetization curves, i.e., the beginning of shielding due to superconducting fluctuations (see Fig.~\ref{fig1}C). The two data points (gray triangles) determined via field sweeps of the DC-magnetization between 3.6\,mK and 6.0\,mK, see Fig.~S4 of SOM, are most likely not related to these A-phase fluctuations; their origin should be addressed by future work. The A + SC phase represents the concurring (dominantly) nuclear AF order and superconductivity, at least at fields below 3 -- 4\,mT. Only at $B = 0$ almost full shielding is observed. The low-temperature limit of our experiment is around 800\,$\mu$K, therefore we cannot detect the fc DC-$M(T)$ peaks above 23\,mT. The two red dashed lines mark the range within which the A-phase boundary line may end. At low fields, $B <$ 2\,mT, a splitting of the transition around 2\,mK in two parts exists, cf. Fig.~S3. The green circle indicates the superconducting transition temperature seen in the AC-susceptibility at $B = 0$ (cf. Fig.~\ref{fig1}D) while the yellow circles (partially covered by the green point) result from the shielding signals in the zero-field-cooled DC-magnetization, see Fig.~\ref{fig1}C. In the inset, these shielding transitions are shown separately on an enlarged scale. As is evident from this, the superconducting phase boundary $T_{\rm c}$ vs. $B$ is extremely steep at low fields, with $-dB_{\textrm{c2}}/dT|_{\rm {T_c}} \approx 25$ T/K. This is of the same gigantic size as found for the canonical heavy-electron superconductor CeCu$_2$Si$_2$ (see Sec.~A, SOM). If superconductivity exists at higher fields, $B_{\textrm{c2}}(T)$ extrapolates to 30 -- 60\,mT (at $T = 0$), i.e., close to the critical field of the primary electronic AF phase.}
\end{center}
\end{figurehere}
\newpage
\begin{figurehere}
\begin{center}
\includegraphics[width=0.9\columnwidth]{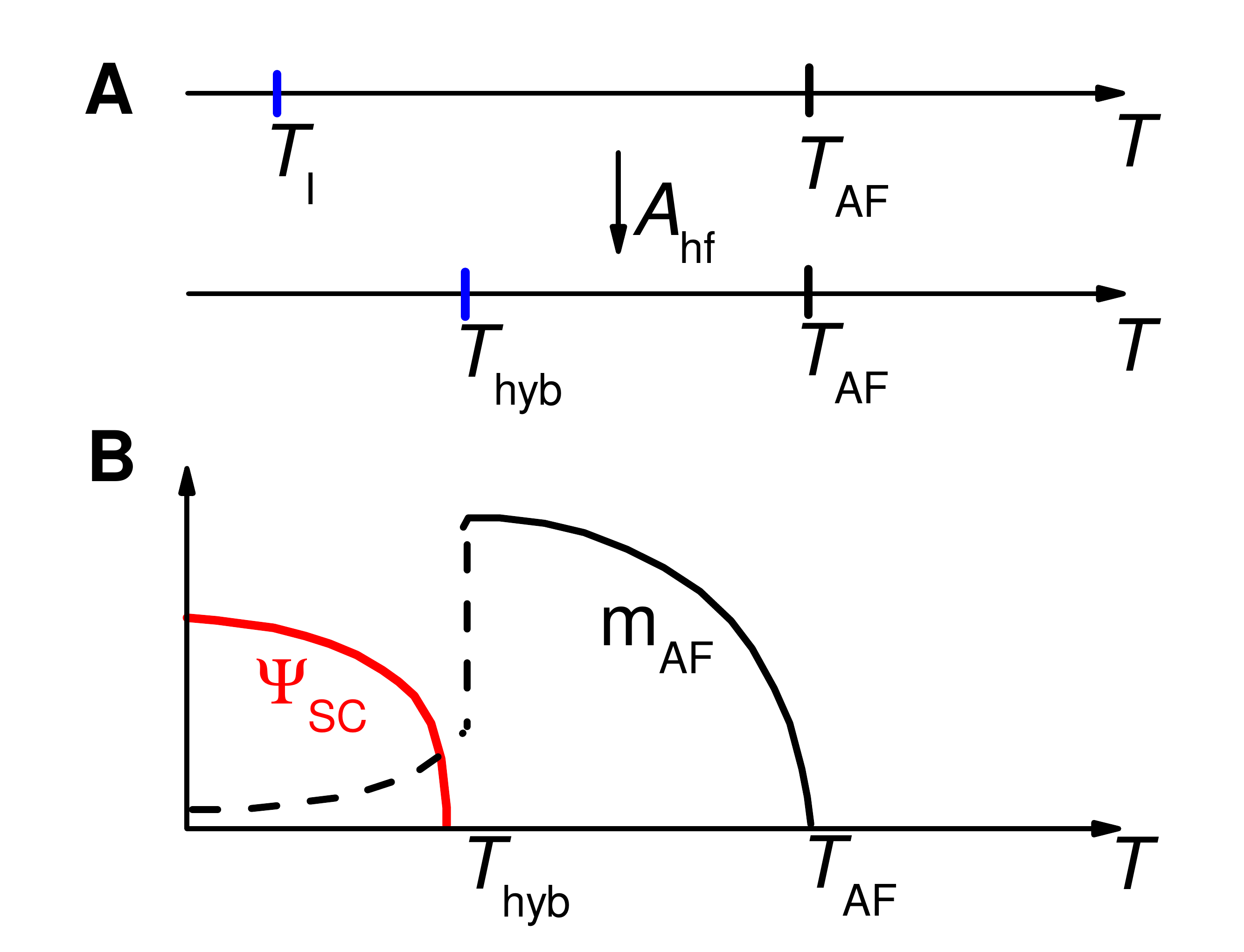}
\caption{\label{fig4}
Sketch of the two phase transitions associated with electronic and nuclear spin orders. (\textbf{A}) Sketch of the two phase transitions associated with electronic and nuclear spin orders. Top line: without a hyperfine coupling ($A_{\textrm{hf}}$), the electronic and nuclear spins are ordered at $T_{\textrm{AF}}$ and $T_{\textrm{I}}$, respectively. Bottom line: with a hyperfine coupling, $T_{\textrm{AF}}$ is not affected, but a hybrid nuclear and electronic spin order is induced at $T_{\textrm{hyb}} \gg T_{\textrm{I}}$. (\textbf{B}) Temperature evolution of the primary electronic spin order parameter ($m_{\textrm{AF}}$) and the superconducting order parameter $\phi_{\textrm{SC}}$. $\phi_{\textrm{SC}}$ is developed when $m_{\textrm{AF}}$ is suppressed by the formation of hybrid nuclear and electronic spin order right below $T_{\textrm{hyb}}$.}
\end{center}
\end{figurehere}
\renewcommand\figurename{FIG. S$\!\!$}
\renewcommand{\theequation}{S\arabic{equation}}
\setcounter{figure}{0}
\newpage
\begin{center}
	{\large\bf Supporting Online Materials for\\ "Emergence of superconductivity in the canonical heavy-electron metal \YRS "}\\
	[0.5cm]
	Erwin Schuberth$^{1,2\,*}$, Marc Tippmann$^{1}$, Lucia Steinke $^{1,2}$, Stefan Lausberg$^{2}$, 
	Alexander Steppke${^2}$, Manuel Brando$^{2}$, Cornelius Krellner$^{2,3}$, 
	Christoph Geibel$^{2}$, Rong Yu$^{4}$, Qimiao Si$^{5\,*}$, and Frank Steglich$^{2,6,7\,*}$\\
	{\textit{$^{1}$Walther Meissner Institute for Low Temperature Research,
			Bavarian Academy of Sciences, 85748 Garching, Germany}\\
		\textit{$^{2}$Max Planck Institute for Chemical Physics of Solids,
			01187~Dresden, Germany}\\
		\textit{$^{3}$Physics Institute,  University of Frankfurt, 
			60438 Frankfurt, Germany}\\
		\textit{$^{4}$Department of Physics, Renmin University of China, Beijing 100872, China.}\\
		\textit{$^{5}$Department of Physics and Astronomy, Rice University,
			Houston, TX 77005, USA}\\
		\textit{$^{6}$Center for Correlated Matter, Zhejiang University, Hangzhou, Zhejiang 310058, China}\\
		\textit{$^{7}$Institute of Physics, Chinese Academy of Sciences, Beijing 100190, China}\\
	}
\end{center}
\vspace{0.8cm}
\subsection{Introductory remarks}
\label{remarks}
\nocite{cuprates,Loehneysen07,Gegenwart,Mathur,Stockert,Custers,Shishido,Park,Knebel08,Loehneysen,Schroeder,Paschen,Friedemann,Si01,Coleman,Senthil,SOM,Schuberth13,Steglich,Steppke,Ronnow,Andres,Steinke,Sichelschmidt,Knebel06}
\YRS\ is a canonical quantum critical heavy-electron compound (see, e.g., Ref.~\onlinecite{Si2010} and references therein). Isothermal magnetotransport~\cite{Paschen,Friedemann} and thermodynamic~\cite{Gegenwart07} measurements have provided evidence that the AF QCP is unconventional, involving a discontinuous jump of the Fermi surface. The finite-temperature signature of this abrupt Fermi surface transformation is a crossover line $T^*(B)$ in the $T - B$ phase diagram. On the elevated-field side of the $T^*(B)$ line, Kondo entanglement between the spins of the $4f$-shells of the Yb$^{3+}$ ions and the spins of the conduction electrons, giving rise to composite heavy electrons as the charge carriers. The breakup of the Kondo entanglement as the field is reduced across the $T^*$ line characterizes the unconventional QCP~\cite{Si01,Coleman,Senthil}. In the main text we have presented an investigation of this material in a temperature range never accessed before. By means of magnetization, susceptibility and heat capacity measurements, we could identify a rich $T - B$ phase diagram (Fig.~3 of the main text), which is characterized by two principal phases: an AF phase below \TN\ = 70\,mK with Yb-derived ordered moments of the order of $10^{-3}\mu_{\textrm{B}}$ (cf. Ref.~\onlinecite{Ishida03}) and a critical field of 60\,mT, when applied within the basal tetragonal (``easy magnetic'') plane and an A + SC phase below \TA\ $\gtrsim 2$\,mK which includes superconductivity. In addition, there is evidence for strong AF (A-phase) fluctuations below \TB\ $\simeq$ 10\,mK. While the AF phase is of purely electronic nature, the new A phase involves a coupling between the magnetic moments due to the Yb-derived $4f$ electrons and the Yb nuclear moments. Because of the size of the critical field \BA\ at which \TA\ $\rightarrow 0$ (30 -- 60\,mT), the A phase must be dominantly of nuclear origin.

While the superconducting phase transition at \TC\ = 2\,mK $\lesssim$ \TA\ is characterized by strong diamagnetic signals in the zfc DC-magnetization (Fig.~1C, main text) and AC-susceptibility (Fig.~1D, main text) as well as a Meissner effect (Figs.~1A and 1B, main text), the huge jump anomaly in the specific-heat coefficient ($\approx$ 1000\,J/K$^{2}$mol) typifies an AF phase transition involving dominating nuclear spins. Importantly, there is a large initial slope (at $T \lesssim$ \TC) of the upper critical field curve ($\approx 25$\,T/K), which points to an effective charge-carrier mass of several $100\,m_{e}$, implying heavy-electron superconductivity associated with the Yb-derived $4f$ electrons~\cite{Steglich}. With transition temperatures in the range of a Kelvin, the upper critical field $B_{\textrm{c2}}(T \rightarrow 0)$ can reach several Tesla and is often controlled by Pauli limiting (see, e.g., Ref.~\onlinecite{Rauchschwalbe} and references therein). For instance, in CeCu$_{2}$Si$_{2}$ \TC\, $\approx$ 0.6\,K and $B_{\textrm{c2}} \approx 2$\,T (Ref.~\onlinecite{Rauchschwalbe}), i.e., $B_{\textrm{c2}}/T_{c} (\textrm{T/K}) \approx 3$, or in UBe$_{13}$ \TC\ $\approx$ 0.8\,K (Ref.~\onlinecite{Ott}) and $B_{\textrm{c2}} \approx 12$\,T (Ref.~\onlinecite{Langhammer}), i.e., $B_{\textrm{c2}}/T_{c} (\textrm{T/K}) \approx 15$. The initial slope of $B_{\textrm{c2}}(T)$ at $T_{\textrm{c}}$ is generically huge for heavy-electron superconductors, e.g., 23 T/K for CeCu$_2$Si$_2$ (Ref.~\onlinecite{Assmus}) and nearly infinite for UBe$_{13}$ (Ref.~\onlinecite{Rauchschwalbe87}).  
\subsection{Cryostat, samples, and thermometry}
\begin{figure}[!ht]
	{\includegraphics[width=0.5\columnwidth]{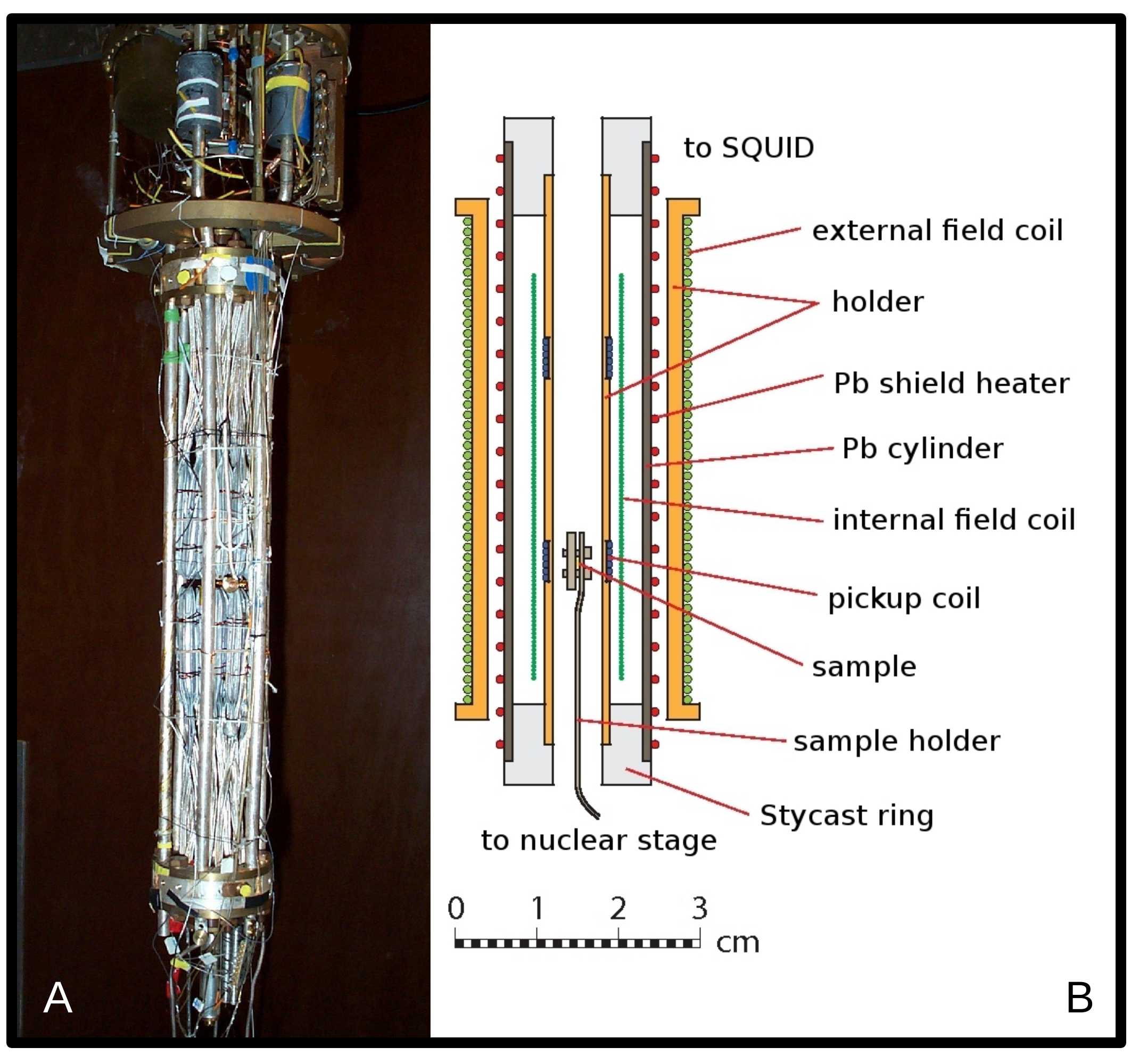}}
	\caption{\label{figS1} (\textbf{A}) Nuclear stage of a home-made demagnetization cryostat at the Walther Meissner Institute. The nuclear stage consists of 0.9 moles of PrNi$_5$ soldered to silver with cadmium flux. The lowest final temperature is 0.4\,mK. The lead shielded cylinders seen on top are two superconducting heat-switches of special design~\cite{Schuberth84}. The uppermost plate is the bottom of the mixing chamber of a home-made dilution refrigerator with a base temperature of 5\,mK. After demagnetizing, the nuclear stage stays below 1\,mK for 2 weeks which implies a residual heat leak of about 6\,nW. (\textbf{B}) SQUID magnetometer used for DC-magnetization and AC-susceptibility measurements. The measuring fields are frozen in the lead cylinder by applying an external magnetic field generated by the outer field coil and by heating the system to a temperature between \TC\ = 7.2\,K (Pb) and \TC\ = 9.2\,K (NbTi), the material of the outer field coil. By cooling subsequently the magnetometer to below \TC\ (Pb) the applied field is frozen-in and stabilized by the superconducting lead cylinder. The available field range is up to 60\,mT, just below the critical field of Pb. The induced current in the gradometer-type pickup system is transferred to an rf SQUID (at 4\,K) through a superconducting flux transformer. The single layer inner coil allows to apply a small magnetic field, e.g., for reducing the residual earth field (vertical component) for zero-field cooled measurements and was also used for field sweeps and for AC-susceptibility experiments. The sample is thermally strongly connected to the nuclear stage by a silver wire but is isolated from the magnetometer which by itself is thermally stabilized at about 20\,mK to avoid variations of the background signal due to thermal drifts.}
\end{figure} 
We have carried out a series of different experiments in the ultra-low-temperature cryostat at the Walther Meissner Institute in Garching which consists of a 0.9 mole PrNi$_{5}$ nuclear demagnetization stage with a final temperature of 0.4\,mK. This is shown in Fig.~S\ref{figS1}A. The temperature of the nuclear stage is determined by pulsed nuclear magnetic resonance on $^{63}$Cu and $^{195}$Pt nuclei. Samples of these metals were screwed to the nuclear stage. This method results in a temperature accuracy of 2 -- 3\,\% even down to 0.8\,mK (Ref.~\onlinecite{Schuberth84}). \YRS\, single-crystalline platelets of very high  quality have been grown in indium flux; careful checks did not reveal any In inclusions. Samples \#1 (2.22 mg) and \#2 (0.75 mg) were tiny, with dimensions of about $2 \times 1 \times 0.08$\,mm$^3$. Their residual resistivity ratios (RRR) were about 150, among the best ever obtained\cite{Krellner}. Samples \#3 (30.62 mg), \#4 (7.1 mg) and \#5 (5.44 mg) were bigger ($\approx$ 30\,mg) with RRR $\approx$ 50; they showed identical results, in particular the same superconducting transition temperature \TC\ (= 2\,mK) was found for all samples studied. This insensitivity of \TC\ on disorder is most likely due to the short superconducting coherence length, which corresponds to the huge effective mass of the charge carriers, i.e., several hundred times the bare electron mass. The \YRS\ samples were clamped on a 5N silver rod which by itself was screwed to the nuclear stage. Thin single crystalline platelets of \YRS\ were placed in the pick-up coil of a superconducting flux transformer, with their flat surface (their basal tetragonal plane) aligned parallel to the axis of the pick-up coil, which transfers the signal to a radiofrequency (rf) Nb-SQUID, see Fig.~S1B. Since the electrical conductivity of the samples is high (of the order of 5000\,S) we have no indication that, after thermalization, their temperature deviated from that of the nuclear stage by more than a few tens of $\mu$K, even at the lowest temperatures. The data were always taken starting several hours after the end of the demagnetization process to assure that the nuclear stage and the samples had reached their final equilibrium temperature.
\subsection {DC-magnetization}
\begin{figure}[t]
	\includegraphics[width=0.7\columnwidth]{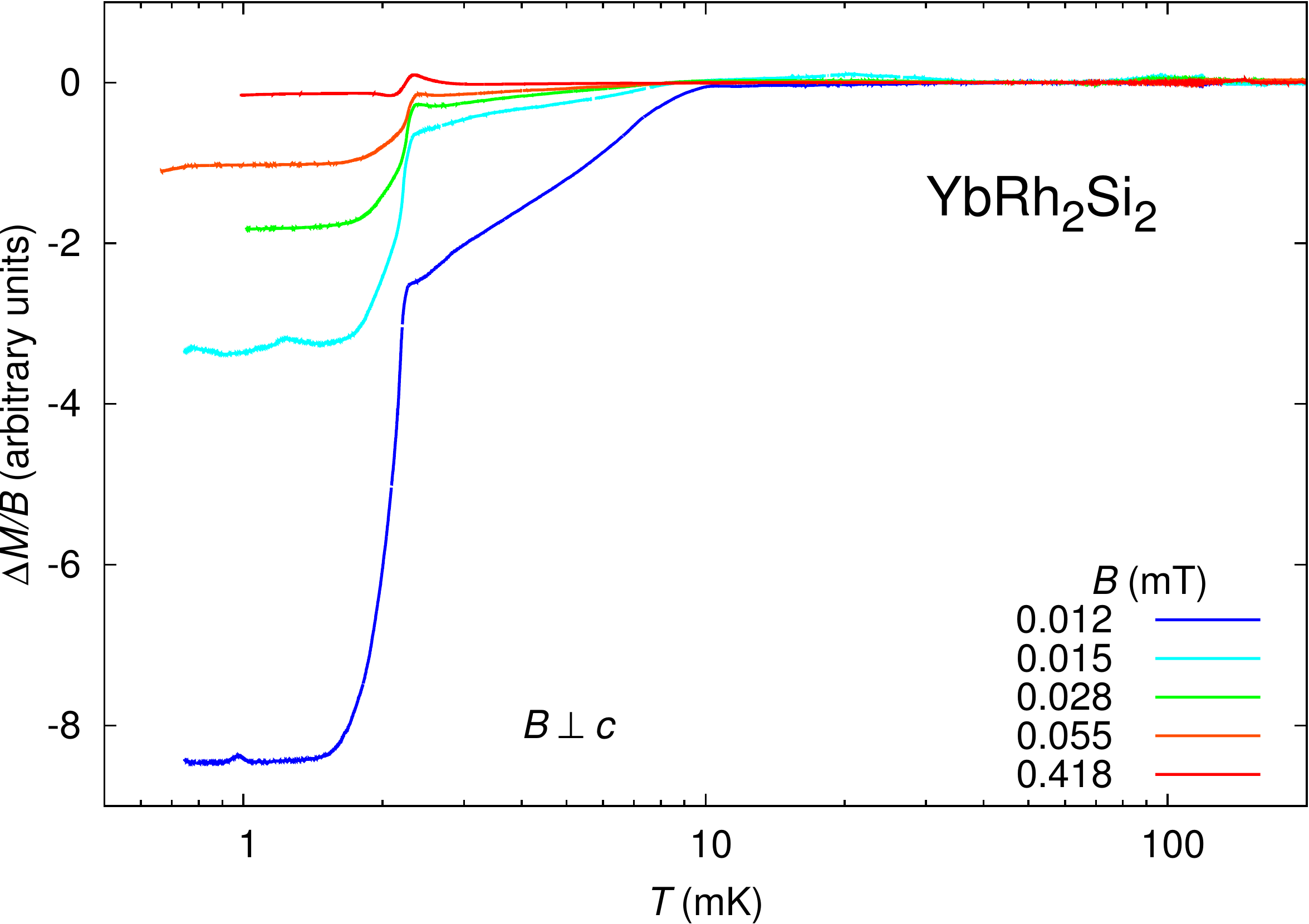}
	\caption{\label{figS2} Differences between zfc and fc DC-magnetization curves of \YRS. The point where the curves meet (\TB\ $\simeq 10$\,mK) is taken as the onset of the "B phase". The sharp drop of $M(T)/T$ at \TC\ = 2\,mK indicates the onset of superconductivity. Measurements were performed on sample \#3.}
\end{figure}
We have measured the DC-magnetization in magnetic fields up to 60\,mT and down to 1\,mK using a home made rf SQUID magnetometer. The main magnetic field $B$ at the sample is provided by a superconducting lead cylinder in which the (vertical) magnetic field generated by an external coil (aligned parallel to the lead cylinder) is frozen-in and stabilized by cooling to below the superconducting transition temperature of Pb. Then, the external coil can be switched off. This way, the measuring field was applied within the basal tetragonal plane of the \YRS\ single crystals. To obtain the smallest possible magnetic fields for zero-field cooled (zfc) measurements, an additional coil around the pick-up coils of the device (placed inside the lead cylinder) was added. The magnetic field therein necessary to compensate the residual earth field plus remaining static fields in the surroundings was determined by the point where the DC-magnetization traces change sign. According to the international geomagnetic reference field (IGRF), the vertical component of the earth field at the location of the laboratory is -0.0433\,mT, which was compensated in almost all our experiments. Although for the set-up and its environment non-magnetic material was used, there were magnetic fields of the order of 0.06\,mT at the position of the sample. After compensation, the vertical component of the measuring field could be reduced to 0.012\,mT (cf. Fig.~\ref{figS1}C of the main text). We note that the horizontal component of the earth field, -20.99\,$\mu$T, does not have to be taken into account: If aligned within the (a,b)-plane of the single crystal, only about 9\,\% of its strength adds to the applied external field, due to the large magnetic anisotropy (Ref.~\onlinecite{Custers}). This contribution is even smaller, when the horizontal earth-field component is not aligned parallel to the (a,b)-plane of the crystal. On the other hand, the highest fields for measuring the zfc traces (Fig.~S2) was limited by the internal field coil to 0.5\,mT.

Following the afore-mentioned procedure, it was possible to measure the zfc and the field cooled (fc) DC-magnetization. Selected traces are shown in Figs.~\ref{figS1}A and~\ref{figS1}B and discussed in the main text. It is worth comparing the zfc and fc results by building the difference between them, as shown in Fig.~S\ref{figS2}. The temperature \TB\ where the curves meet is taken as the onset of the fluctuations of the A phase, resulting in partial shielding due to superconducting fluctuations (see below). The sharp drop of zfc $M(T)/B$ at \TC\ = 2\,mK indicates the transition into a coherent superconducting phase. 
\begin{figure}[b]
	{\includegraphics[width=0.9\columnwidth]{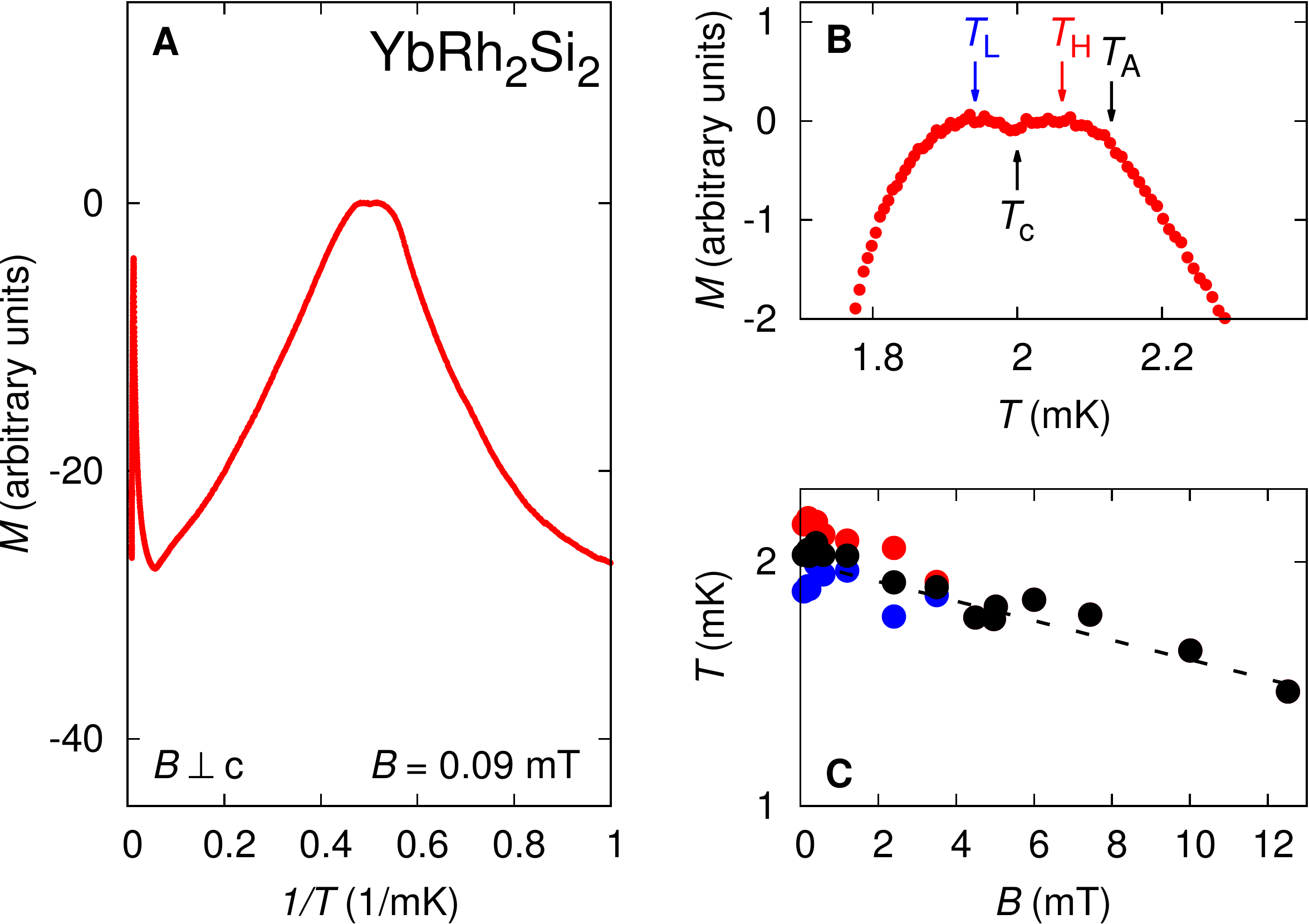}}
	\label{figS3}
\end{figure}
\begin{figure}[!ht]
	\caption{(\textbf{A}) Field-cooled DC-magnetization taken at a very low field of $B$ = 0.09\,mT plotted as a function of $1/T$. The magnetization shows a plateau of about 0.2\,mK around 2\,mK which indicates a separation between the transition temperatures \TA\ and \TC. (\textbf{B}) The same magnetization plotted as a function of $T$ in the vicinity of \TC\ = 2\,mK; \TA\ is determined by the temperature where $M(T)$ changes slope at the high-field side, and \TC\ has been determined by shielding experiments, i.e., of the AC-susceptibility and zfc DC-magnetization, see main text. Here all features of these phase transitions can be seen in detail: The increase of $M(T)$ upon cooling at the high-field side reflects the decrease of the staggered magnetization of the primary order due to the competing A-phase fluctuations below \TB\ $\approx 10$\,mK. The broad maximum at $T_{\textrm{H}}$ (red arrow) results from an overcompensation of the increase in $M(T)$, i.e., the weakening of the primary order because of the competing nuclear order, by the reduction of the magnetization upon the onset of the small (2\,\%) electronic component, $\phi_J$, of the hybrid order at \TA. Just at \TC, another increase of the fc DC-magnetization points to an additional reduction of the staggered magnetization, which appears to be related to the superconducting phase transition being of first order (see Fig.\,S6). Finally at $T_{\textrm{L}}$ (blue arrow), slightly below \TC, $M(T)$ decreases again due to the expulsion of magnetic flux. (\textbf{C}) The peak positions $T_{\textrm{H}}$ and $T_{\textrm{L}}$ (from \textbf{B}) as well as their mean values are shown by the red, blue and black symbols as a function of the magnetic field. This illustrates the separation of \TA\ (slightly above $T_{\textrm{H}}$) and \TC\ (almost equal to the mean value ($T_{\textrm{H}}$ + $T_{\textrm{L}}$)/2) below 3 -- 4\,mT. The field dependence of \TC\ is in excellent agreement with the inverse of the initial slope of $B_{\textrm{c2}}(T)$ at $B = 0$ (dashed line), i.e., $\approx 25$\,T/K (cf. inset of Fig.~3, main text). The error in temperature is of the order of $\pm 0.1$\,mK (5\,\%). It results from the uncertainty of the positions of $T_{\textrm{H}}$ and $T_{\textrm{L}}$ and from the calibration of the temperature of the nuclear stage. Measurements were performed on sample \#1.}
\end{figure}
The pronounced signal below 2\,mK means substantial shielding of the external magnetic field. Furthermore, this shielding still occurs at \TC\ $\approx 2$\,mK up to 0.418\,mT, which implies that the initial slope of the upper-critical-field curve is extremely large and that, presumably, superconductivity extends to much higher magnetic fields, as is concluded from the peaks in the temperature dependence of the fc DC-magnetization which indicates the onset of the Meissner effect (cf. Fig.~1B, main text). The Meissner volume is obtained from the ratio of the decrease in fc DC-magnetization upon cooling through \TC\ and the shielding signal in the zfc $M(T)$. As inferred from Fig.~1C (main text), the Meissner volume amounts to less than 3\,\% at the lowest field of 0.012\,mT. Such a small value is in line with what is commonly observed for \textit{bulk} type-II superconducting samples. For instance, bulk polycrystalline CeCu$_{2}$Si$_{2}$ showed a Meissner volume of not more than $\approx 2$\,\%. However, after the pinning centers were significantly removed by powdering the polycrystals and subsequently annealing the powder, the Meissner volume turned out to increase to $\approx 70$\,\% (Ref.~\onlinecite{Rauchschwalbe}).

%
%
In Fig.~S3A we have plotted the fc DC-magnetization taken at $B$ = 0.09\,mT as a function of $1/T$. It is interesting to note that $M$ vs $1/T$ shows a kind of plateau of about 0.2\,mK around 2\,mK, while $M$ vs $T$ exhibits a faint double-peak structure (Fig.~S3B). This indicates a separation between \TA\ and the superconducting transition \TC. Furthermore, we find that at about 3 -- 4\,mT, both phase transitions merge within the experimental resolution, see Fig.~S3C.

In addition to the phase-transition and cross-over anomalies shown at a very low field in Fig.~\ref{figS1}A of the main text and as already discussed in Sec.~A, we observe an increase of $M(T)/B$ upon cooling to below \TB\ $\simeq 10$\,mK down to \TC\ where the data reveal a sharp transition. This increase of the low-field magnetization within the AF phase can be described by a Curie Weiss law. The fits around 60\,mT yield Weiss temperatures very close to zero, suggesting nearly free paramagnetic moments of about 1.45\,$\mu_{\textrm{B}}$ for temperatures on the order of 25\,mK and 0.1\,$\mu_{\textrm{B}}$ for temperatures on the order of 2.2\,mK. In addition, $M$ vs $T$ at $B \approx 60$\,mT is well fitted by a Brillouin function. Using the so-derived fitting parameters as a guidance for the magnetization at elevated fields, say close to 50\,mT, we can infer that the differential magnetic susceptibility decreases upon cooling at temperatures 
on the order of the 10\,mK. This behavior is to be contrasted with the upturn of the magnetic susceptibility near \TB\ in the low-field linear-response regime (where $M/B$ is identical to the magnetic susceptibility), as shown in Fig.~1A of the main text.
\begin{figure}[t]
	\hspace{-3cm}
	{\includegraphics[width=0.9\columnwidth]{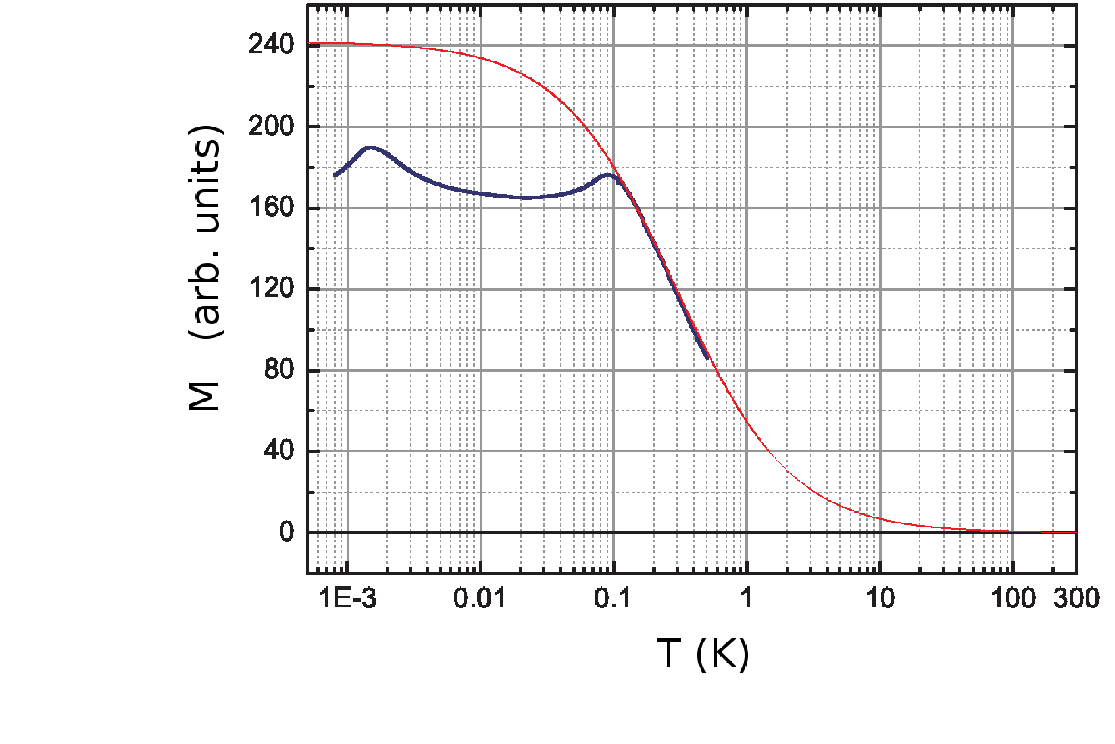}}
	\caption{\label{figS10} Curie-Weiss fit to the DC magnetization curve in a magnetic field of 10.1\,mT. Scaling with known magnetic units yielded a magnetic saturation moment of 1.24\,$\mu_{\textrm{B}}$. The drop of the magnetization below 2\,mK corresponds to at least 6\% of the saturation magnetization as it seems to further decrease at lower temperatures. As stated in the text, the contribution from the $4f$-electronic part of the nuclear-dominated hybrid order has an upper limit of 1/3 of the drop, 2/3 are due to the Meissner effect. The data were taken from sample \#2.}
\end{figure} 

We interpret this upturn in the uniform magnetic susceptibility at low fields as evidence for the initial decrease of the staggered magnetization associated with the primary electronic AF order. The latter appears to be caused by the development of the fluctuating (nuclear-dominated) hybrid order near \TB, as evidenced by the gradual onset of the nuclear-spin entropy at temperatures just below \TB\ (Fig.~2C in the main text). This is also consistent with the observation of superconducting fluctuations below \TB, which we infer from zfc DC-magnetization (Fig.~1C of the main text) as well as AC-susceptibility in the low-field regime (Fig.~1D of the main text). Upon further decreasing the temperature towards \TA, an actual phase transition into the hybrid order takes place. The associated onset of its small ($\leq 2$\,\%) $4f$-electronic component, the $\phi_J$ order, yields a decrease in the uniform magnetic susceptibility, which over-compensates the tendency of an increasing uniform susceptibility due to a reduction of the primary order, caused by the competing nuclear order. This is inferred from the flattening and subsequent decrease of the susceptibility below \TA, see Fig.~S3B.
\begin{figure}[b]
	{\includegraphics[width=0.7\columnwidth]{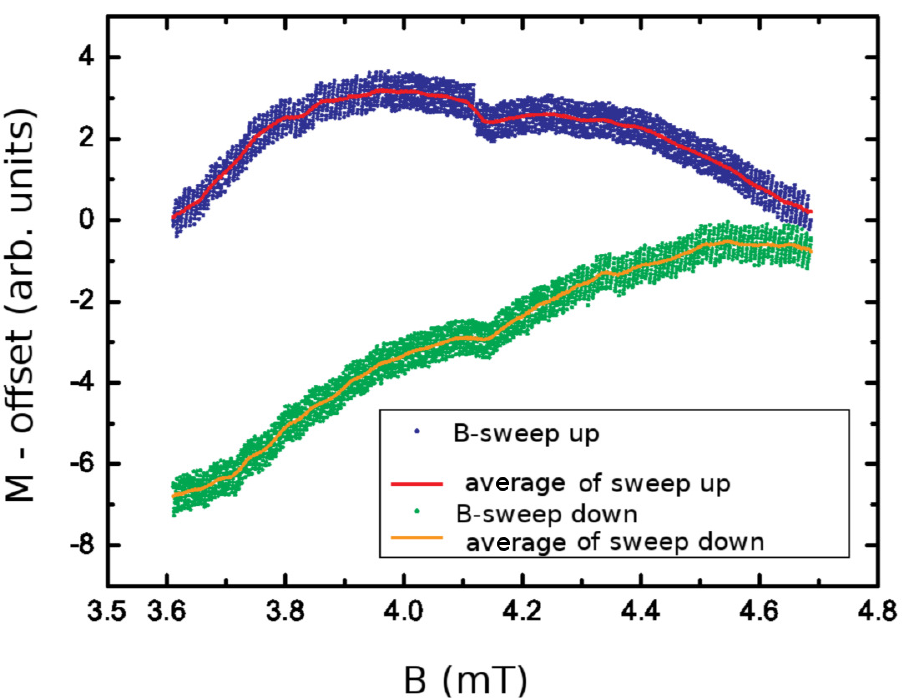}}\\
	\caption{\label{figS6} $B$-sweep around 4\,mT. To get more information about the phase diagram, the external field in one experiment was swept slowly upwards and downwards at a constant temperature of 5\,mK. The data shown here represent the static magnetic susceptibility during this sweep. Since our astatic pick up coils were not perfectly matched, a steep increase, here denoted offset, was subtracted from the raw data. The apparent hysteresis is largely due to changes in this offset. The origin of the observed kinks remains unclear. Measurements were performed on sample \#3.}
\end{figure}
As also shown in this figure, there is another increase in the susceptibility upon further cooling, which we ascribe to the first-order nature of the superconducting phase transition, see Fig.~S\ref{figS7}. However, this rise eventually becomes overcompensated by the Meissner effect. As already mentioned, the separation between \TC\ and \TA\ is visible at only low fields, up to about 3 -- 4\,mT. At $B = 2.4$\,mT, where the specific-heat results yield \TA\ $\simeq 2$\,mK (Fig.~2B, main text), the splitting between \TA\ and \TC\ is still visible in the magnetization: In Fig.~S3C, we show the positions $T_{\rm H}$ and $T_{\rm L}$ of the two sub-peaks as well as that of the mean value $(T_{\rm H} + T_{\rm L})/2$ which is almost identical with \TC\ ($\simeq 1.9$\,mK at $B = 2.4$\,mT). The field dependence of these mean values below 3--4\,mT determines the initial slope of the upper critical field, $B_{\rm c2}' = -dB_{\rm c2}(T)/dT$ (at \TC) $\simeq 25$\,T/K. It is reassuring to see that this value derived from the fc DC-magnetization (Meissner effect) agrees well with $B_{\rm c2}'$ determined from the zfc DC-magnetization measurements (shielding effect), see inset of Fig.~3, main text. This is a value typical for heavy-electron superconductivity such as the one in CeCu$_{2}$Si$_{2}$ (Ref.~\onlinecite{Assmus}).
\begin{figure}[t]
	{\includegraphics[width=0.7\columnwidth]{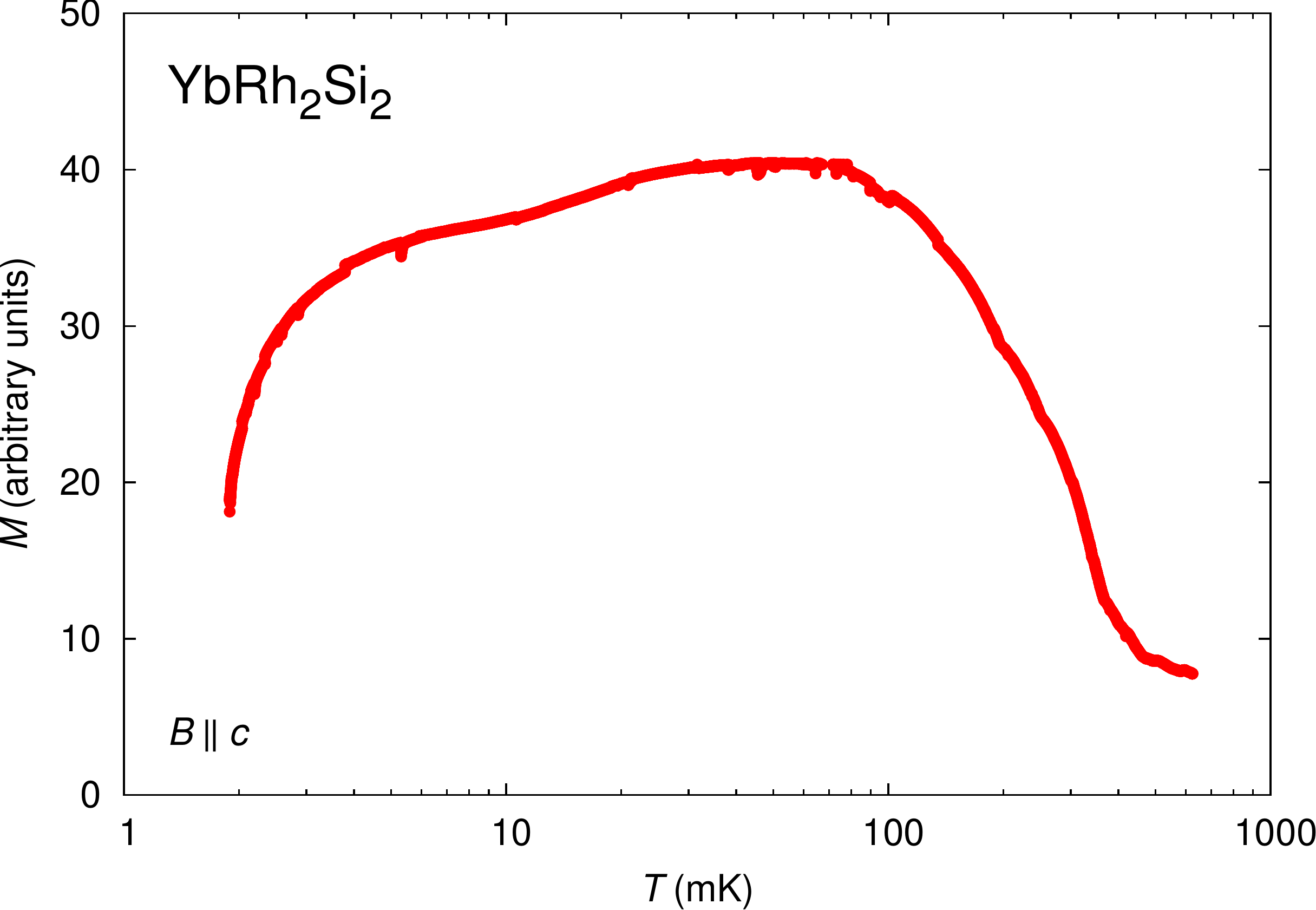}}
	\caption{\label{figS5} Field-cooled DC-magnetization of \YRS\ at $B = 0.7$\,mT, with the magnetic field $B$ aligned parallel to the $c$-axis. The three features observed with $B \perp c$ can be also observed here: the AF transition at \TN\ = 70\,mK, the onset of the B phase at \TB\ = 10\,mK and the sharp drop into the superconducting phase at \TC\ $\approx$ 2\,mK. Measurements were performed on sample \#2.}
\end{figure} 

A quantitative analysis of the fc $M(T)$, using the data shown in Fig.~S\ref{figS10} obtained at $B = 10.1$\,mT in the temperature range 0.8\,mK $\leq T \leq  540$\,mK, reveals a saturation moment of 1.24\,$\mu_{\rm B}$. The $M(T)$ decrease on the low-temperature side of the peak at 2\,mK as measured amounts to $\simeq 0.075$\,$\mu_{\rm B}$. This is a lower limit as the magnetization will certainly further decrease upon further cooling. The electronic $\phi_{J}$ component of the hybrid A-phase will contribute to this reduction of $M(T)$ less than 0.025\,$\mu_{\rm B}$, i.e., at best 1/3. As the nuclear order cannot be resolved in our magnetization measurements because of the small nuclear moment, at least 2/3 of the $M(T)$ decrease \textit{must} be due to the Meissner effect. This strongly supports our argument that superconductivity coexists with the nuclear dominated hybrid A-phase at fields larger than 3 -- 4\,mT, where no separation between \TA\ and \TC\ 
can be resolved anymore (see main text).

To further explore the $B - T$ phase diagram of \YRS\ we performed field-sweep experiments between 3 and 5\,mT while trying to keep the temperature constant ($T$ varied between 4 and 5\,mK), see Fig.~S\ref{figS6}. We observe two distinct kinks at about 4.2\,mK. The positions of the observed kinks are indicated by the gray triangles in Fig.~3 of the main text. Future experiments are necessary to resolve the origin of these kinks.

Motivated by the discovery that \YRS\, under chemical pressure shows ferromagnetic (FM) order with moments oriented suprisingly along the crystallographic c-axis, i.e., the magnetic hard axis~\cite{Lausberg}, we performed a few magnetization measurements with $B \parallel$ c. Because of the strong crystalline anisotropy the magnetization along the c-axis is smaller by a factor of about 11 than that in the basal plane. Therefore, we had to apply a large magnetic field to detect a signal. The results of a measurement with $B = 0.7$\,mT are shown in Fig.~S\ref{figS5}. The three features observed with $B \perp$ c can be also observed here: the AF transition at \TN\ = 70\,mK, the onset of the B phase at \TB\, = 10\,mK and the sharp drop into the A + SC phase at \TC\ = 2\,mK. This definitely rules out a FM transition with moments along the c-axis at \TA\ ($\gtrsim$ \TC) to exist at ambient pressure. 

%
%
\subsection{AC-susceptibility}
\begin{figure}[t]
	{\includegraphics[width=0.7\columnwidth]{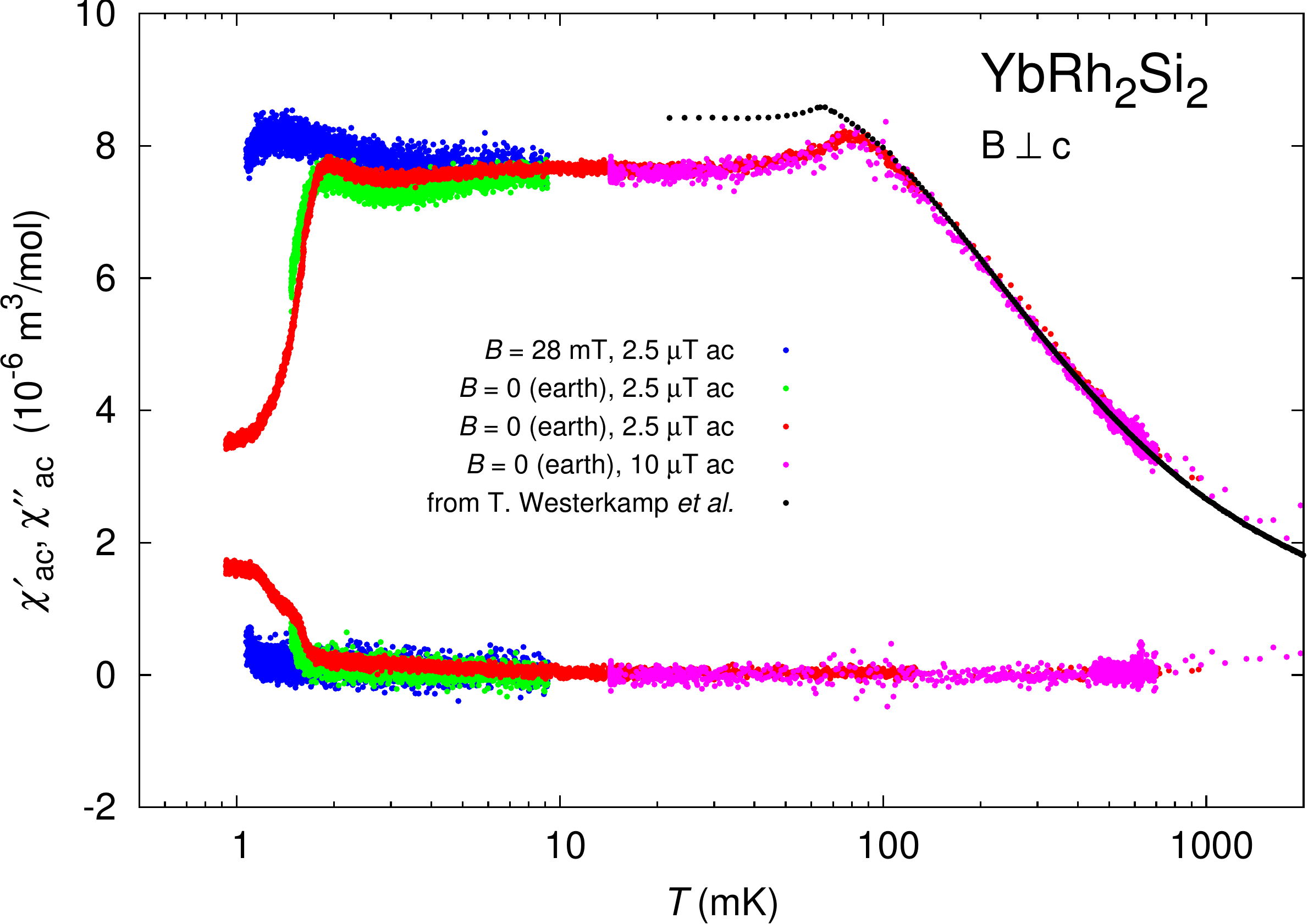}}
	\caption{\label{figS7} Temperature sweeps of the AC-susceptibility of \YRS\, measured with a conventional mutual inductance setup in earth magnetic field and earth magnetic field plus 28\,mT. The in-phase  $\chi'_{\textrm{ac}}(T)$ and out-of-phase  $\chi''_{\textrm{ac}}(T)$ responses were measured with a two-channel lock-in amplifier. With this setup it was possible to measure temperature sweeps from 1\,mK up to about 1000\,mK and scale the data with those taken in a standard Kelvinox 400 dilution refrigerator (Oxford Instruments) down to 20\,mK (black points)~\cite{Westerkamp}. We used an excitation field of 2.5\,$\mu$T and a frequency of 117\,Hz. With higher excitation fields it was not possible to cool the sample down to 1\,mK. For instance, with 10\,$\mu$T the lowest temperature was about 10\,mK (magenta points). Signatures of the superconducting transition and the crossover into a regime with A-phase fluctuations as well as of the electronic AF transition are clearly visible. Most importantly, at \TC\, = 2\,mK the imaginary part of the susceptibility shows a clear increase, suggesting a first order nature of the transition. Measurements were performed on sample \#5.}
\end{figure} 
The AC-susceptibility $\chi'_{\textrm{ac}}(T)$ was measured using two methods: 1) by modulating the rf SQUID system with frequencies between 17 and 87\,Hz (here the earth field could be compensated as in the DC case) and 2) by a conventional mutual inductance setup in the center of a big magnet which was also thermally  connected to the nuclear stage; this allowed frequencies up to a few hundred Hz (here the smallest field was the earth field). Using the SQUID magnetometer with 17\,Hz and in the ``virgin'' state (excitation field between 2 and 5\,nT), negative values of $\chi'_{ac}(T)$ below $T \approx 2$\,mK within the A phase were reproducibly detected, as shown in Fig.~\ref{figS1}D of the main text. Great care was taken to determine the zero of the susceptibility in the limit $T \rightarrow \infty$. Unfortunately, heating the sample to temperatures above 600\,mK resulted in a thermal drift of the signal due to a warm-up of the mixing chamber of the dilution refrigerator, making measurements unreliable at higher temperatures. Zero magnetization was therefore determined by a comparison of the signal with and without the sample under identical conditions. With the conventional mutual inductance setup it was possible to measure $T$-sweeps from 1\,mK up to about 1000\,mK and scale the data with those taken in a standard Kelvinox 400 (Oxford Instruments) down to 20\,mK (black points in Fig.~S\ref{figS7})~\cite{Westerkamp}. The in-phase $\chi'(T)$ and out-of-phase $\chi''(T)$ responses of selected measurements are shown in Fig.~S\ref{figS7}. We used an excitation field of 2.5\,$\mu$T and a frequency of 117\,Hz. With higher excitation fields it was not possible to cool the sample to below 10\,mK (see, e.g., magenta points in Fig.~S\ref{figS7}). Signatures of both the superconducting and the B phase as well as the electronic AF transition are clearly visible. Most importantly, at \TC\, = 2\,mK the imaginary part of the susceptibility shows a clear increase strongly suggesting that the superconducting transition measured in the earth magnetic field is of first order. This resembles the case of A/S-type CeCu$_{2}$Si$_{2}$ (Ref.~\onlinecite{Feyerherm}), implying the absence of microscopic coexistence between electronic AF order (\TN\, = 70\,mK) and superconductivity. With this setup it was not possible to screen the earth magnetic field, and this might be the reason why the superconducting transition is not manifested by negative values like in Fig.~\ref{figS1}D of the main text. We tried to use a $\mu$-metal cylinder to screen the susceptometer, but its large size resulted in too high a thermal load and did not allow us to cool the low-$T$ stage below 2\,mK.
\subsection{Specific heat}
\begin{figure}[t]
	\begin{center}
		{\includegraphics[width=0.7\columnwidth]{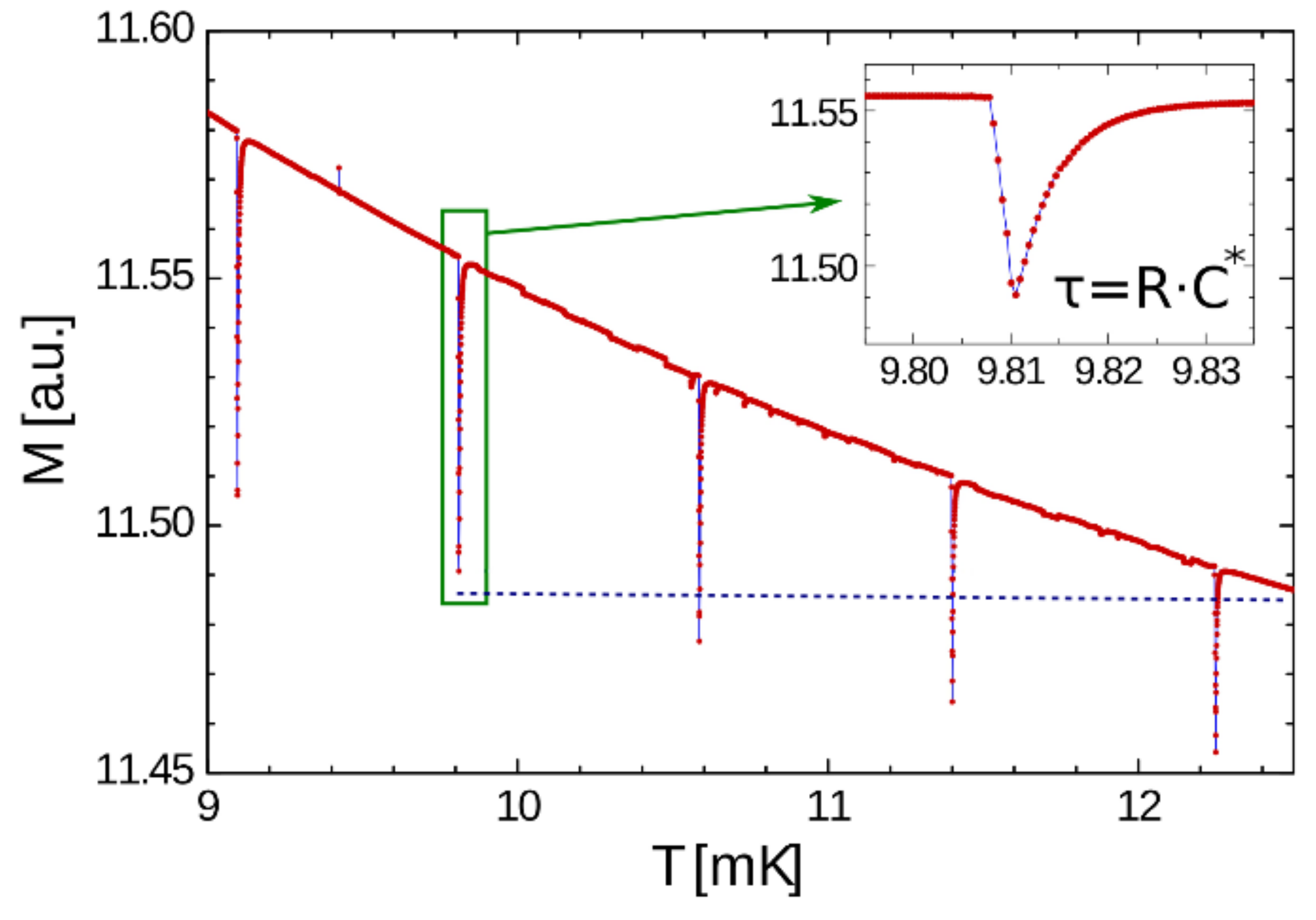}}
	\end{center}
	\caption{\label{figS8} Method for measuring the specific heat of \YRS. During the warm-up, heat pulses of known energy $\Delta Q$ were applied in regular time intervals resulting in an increase of temperature  $\Delta T$. This could be found by first determining the peak (here minimum) of the $M(T)$ curve at the time of the heat pulse (by back-extrapolation of the relaxation curve) and then projecting this value onto the $M(T)$ curve. The total heat capacity value $C^{*}(T)$ is thus obtained from $C^{*}(T) = \Delta Q$/$\Delta T$. In addition, $C^{*}(T)$ can be obtained from the relaxation time of the cooling curve using $\tau = R \cdot C^{*}$ (inset) where $R$ is the thermal resistance between the sample and the nuclear stage. The curve in the figure was taken with $B = 2.4$\,mT. Measurements were performed on sample \#3.}
\end{figure} 
The specific heat of \YRS\, was measured with the semiadiabatic heat-pulse method. Due to its fast thermal response we were able to use the sample itself as a thermometer. The temperature was determined directly from the DC-magnetization, once the temperature dependence of the magnetization was known (see Fig.~S\ref{figS8}). After applying a known heat pulse $\Delta Q$, the temperature increase could be measured by projecting the peak of the pulse onto the warm-up curve (see horizontal dashed line in Fig.~S\ref{figS8}). The heat capacity $C^{*}(T)$ of the sample and addendum is simply obtained from the relation $C^{*}(T) = \Delta Q / \Delta T$. The temperature of the pulse was taken as the mean value between $T_{\textrm{low}}$ and $T_{\textrm{high}}$, the temperatures just before the pulse and at the peak maximum. Simultaneously, $C^{*}(T)$ could be obtained from the relaxation time $\tau$ (determined from the cooling curve) by the relation $\tau = R  \cdot C^{*}$ where $R$ is the thermal resistance of the weak link (high-purity Ag wire) connecting the sample to the nuclear stage. The molar specific heat $C(T)$ of \YRS\ was obtained by subtracting from $C^{*}(T)$ the contribution of the addendum and dividing this difference by the number of moles of the sample. In the temperature range of interest, the thermal conductance $K = 1/R$ of the high-purity Ag wire between the sample and the nuclear stage (divided by the temperature) could be determined to be $K/T \approx (1.06 \pm 0.02)\times 10^{-4}$\,W/K$^{2}$ = const. For example, the relation $C^{*} = K \cdot \tau$ with $C^{*}/T \approx 5$\,J/K$^{2}$mol at 10\,mK yields $\tau \approx 0.25$\,s which is a quite short relaxation time. Around 2\,mK, $C/T$ assumes very large values of the order of 1000\,J/K$^{2}$mol which implies $\tau \approx 50$\,s. This procedure was especially helpful around 2\,mK where the relaxation times were long and the direct temperature reading had large errors because after the heat pulse, the peaks were rounded and their heights (taken as the back extrapolation of the decay curve to the time of the heat pulse) were difficult to determine. The determination of the contribution of the addendum (4\,g of Ag) was difficult as well. It could not be measured separately (we had no fast thermometer for this) but had to be calculated. This resulted in large errors above $T = 10$\,mK. At both, 2.4 and 59.6\,mT, two separate sets of measurements were analyzed in this way, yielding for each field a total of 4 data sets from the two methods. These experimental results are shown in Fig.~\ref{figS2}A of the main text together with a set of theoretical results for the temperature dependence of the nuclear specific heat at selected field-induced ordered magnetic moments. These calculations were performed as described in Ref.~\onlinecite{Steppke}; we have chosen induced moments of 0.05 and 0.15\,$\mu_{\textrm{B}}$/Yb, respectively, so that the calculated curves describe the data measured below 10\,mK at $B = 2.4$\,mT and 59.6\,mT  within the error bars. The 
field-induced increase of the effective moment agrees very well with that derived from the bulk magnetization measured at 50\,mK in the same field range~\cite{Brando}.

As mentioned in the main text, the huge specific heat around \TA\ $\gtrsim 2$\,mK may suggest a nuclear Kondo effect to be operating. On the other hand, with a hyperfine coupling strength of about 25\,mK (see Sec.~G) and an effective Fermi temperature given by the (electronic) Kondo temperature $T_{\rm K} \simeq 25$\,K, the nuclear Kondo temperature is expected to be of order $T_{\rm K}\exp(-1000)$ only. For the nuclear Kondo effect being involved in the Cooper-pair formation at \TC\ = 2\,mK, the nuclear Kondo temperature should be of the order of, at least, 10\TC. In case that the latter would indeed be as large as, say, 25\,mK, the effective quasiparticle mass enhancement would have to be of the order of the bandwidth ($\sim 1$\,eV) over 2.5\,$\mu$eV which amounts to $4 \cdot 10^{5}$. Future theoretical investigations are necessary to check whether such an enormous enhancement of the nuclear Kondo scale can be achieved by suitable renormalizations of coupling constants. As shown in the inset of Fig.~3 (main text) and in Fig.~S3C, the initial slope of the upper critical field vs temperature dependence at \TC\ is approximately 25\,T/K, typical for ordinary heavy-electron superconductivity where the effective mass enhancement varies between 100 and 1000. A mass enhancement of $4 \cdot 10^{5}$, however, should cause an almost vertical initial slope in $B_{\rm c2}(T)$ at \TC. Further experiments to unravel this issue are highly welcome.
\subsection{Hyperfine coupling, nuclear spin entropy and A-phase fluctuations}
As suggested by the observation of a single M\"ossbauer line~\cite{Knebel06}, the nuclear moments in \YRS\ just feel the mean dipolar and quadrupolar fields due to the fast relaxing Yb-derived $4f$ moments (\TK\ = 25\,K), which implies that the hyperfine coupling can be expressed by a $4f$-electron induced hyperfine magnetic field to which the nuclear moments react. The hyperfine coupling is then identical for all Yb isotopes and amounts to 102\,T/$\mu_{\textrm{B}}$ (Refs.~\onlinecite{Kalvius,Shimizu}), as used for the calculation of the nuclear specific heats (Ref.~\onlinecite{Steppke}). These isotopes are characterized by nuclear spin $I = 1/2$, natural abundance 14.3\,\%, nuclear moment 0.488\,$\mu_{\mathrm{N}}$ and hyperfine coupling constant $A = 102$\,T/$\mu_{\mathrm{B}}$ for $^{171}$Yb and 5/2, 16.1\,\%, -0.68\,$\mu_{\mathrm{N}}$, -102\,T/$\mu_{\mathrm{B}}$ for $^{173}$Yb, respectively. The coupling between the $4f$ magnetic moments and the Yb-derived nuclear magnetic moments is ferromagnetic for both isotopes. For $^{103}$Rh in \YRS\ the hyperfine coupling is not known. Observed values for the Knight shift, $K \leq 0.8$\,\%, in a number of Rh-based compounds suggest that the value for Rh metal, 22\,T/$\mu_{\textrm{B}}$ ($K = 0.4$\,\%), is a reasonable approximation~\cite{Seitchik}. For $^{29}$Si, the hyperfine coupling constant obtained from in-plane NMR measurements at fields down to 0.15\,T amounts to -0.073\,T/$\mu_{\textrm{B}}$ (Ref.~\onlinecite{Ishida}). As shown in Fig.~2C (main text), the nuclear spin entropy $S_{\textrm{I}}(T)$ saturates, reaching $S_{\textrm {I,tot}}$, at about $T = 10$\,mK, where the difference between $C(T)/T$, measured at the lowest field of 2.4\,mT, and the nuclear quadrupole contribution ($B = 0$) vanishes within the experimental uncertainty. $S_{\textrm{I,tot}}$ $\simeq 1.35\textrm{R}\ln 2$ consists of the contributions of $^{29}$Si, $S_{\textrm {Si}}$ = 0.033R, $^{103}$Rh, $S_{\textrm {Rh}}$ = 0.693R, $^{171}$Yb and $^{173}$Yb, $S_{\textrm {Yb,tot}}$ = 0.211R. As $S_{\textrm{Si}}$ and $S_{\textrm {Rh}}$ are temperature independent in the temperature range of interest ($T \geq$ 1mK), the 6\,\% drop of $S_{\textrm {I}}(T)$ when cooling from 10 mK to 2 mK, corresponds to a 26\,\% drop of $S_{\textrm{Yb}}(T)$, while 74\,\% of the entropy of the nuclear Yb spins is released below $T_{\textrm A}$. The substantial drop of $S_{\textrm{Yb}}(T)$ below $T \approx 10$\,mK concurs with a significant increase of the fc DC-magnetization upon cooling at low fields (cf. Fig.~1A, main text) and a partial superconductivity shielding (see Figs.~1C and 1D, main text). We ascribe these observations to fluctuations of the A phase when cooling to $T \approx$ \TB, i.e., way above the formation of the hybrid A-phase order at \TA\ $\gtrsim 2$\,mK, cf. Sec.~C. We wish to note that the large Yb-derived nuclear spin entropy of 26\,\% which is released up to temperatures of order 10\,mK is highly consistent with an (antiferro)magnetic phase transition to take place at \TA\ $\simeq 2$\,mK.

\subsection{Landau theory of the magnetic orders involving both electronic and nuclear spins}
\label{theory}
\subsubsection{Three-component theory}
To understand the observed magnetic transitions at both $T_{\rm AF} = 70$\,mK and \TA\ $\gtrsim 2$\,mK, we propose a minimal Landau theory with three relevant spin components: an electronic AF order parameter $m_{\mathrm{AF}}$ at wavevector $\boldsymbol{Q}_{\mathrm{AF}}$, which accounts for the transition at $T_{\mathrm{AF}}$, and two linearly coupled nuclear and electronic order parameters $m_I$ and $m_J$ at $\boldsymbol{Q}_1\neq\boldsymbol{Q}_{\mathrm{AF}}$,
which are pertinent to the transition at $T_A$. We note that a two-component model involving two linearly coupled order parameters, $m_{\mathrm{AF}}$ and a nuclear spin component $m_N$ (both at wavevector $\boldsymbol{Q}_{\mathrm{AF}}$),
is insufficient to understand these two magnetic transitions. For instance, there would be only one $Z_2$ symmetry in the model, and only one magnetic transition is possible.

In addition to the three-spin components in our minimal Landau model, formally, one should also consider the nuclear spin order parameter $m_N$, which linearly couples to $m_{\mathrm{AF}}$. But if $m_{\mathrm{AF}}$ is the dominant order parameter in the interested temperature range, we may neglect the effect of $m_N$, leaving the three-spin components which we consider.

More microscopically, $m_{\mathrm{AF}}$ and $m_J$ are associated with the Yb 4$f$ electronic spins whereas $m_I$ describes the Yb nuclear spins. We define the normalized dimensionless order parameters $\phi_\mathrm{AF}$, $\phi_J$, and $\phi_I$ via $m_{\mathrm{AF}}=g_{el}\phi_\mathrm{AF}$, $m_J=g_{el}\phi_J$, and $m_I=g_{I}\phi_I$, where $g_{el}\approx1$, and $g_{I}\approx3\times10^{-4}$, are the g-factors of the Yb 4$f$ electronic and nuclear (averaging over the $I=1/2$ $^{171}$Yb and $I=5/2$ $^{173}$Yb isotopes) spins, respectively. As described in the main text, the free energy functional of the three-component Landau theory reads
\begin{eqnarray}\label{FE2}
f &=& \frac{r_\mathrm{AF}}{2} \phi^2_\mathrm{AF} + \frac{u_\mathrm{AF}}{4}  \phi^4_\mathrm{AF} + \frac{r_J}{2} \phi^2_J + \frac{u_J}{4} \phi^4_J \nonumber\\
&& + \frac{r_I}{2} \phi^2_I + \frac{u_I}{4} \phi^4_I - \lambda \phi_J \phi_I \nonumber\\
&& + \frac{\epsilon}{2} \phi^2_\mathrm{AF} \phi^2_I + \frac{\eta}{2} \phi^2_J \phi^2_\mathrm{AF},
\end{eqnarray}
where $r_\alpha=T-T_{\alpha}$,
for $\alpha=\mathrm{AF},\ J,\ I$, describe the quadratic couplings.
Here,  $T_{\alpha}$ is the bare ordering temperature when it is positive, and specifies the excitation gap when it is negative. We take $T_\mathrm{AF}=70$ mK, and assume $T_J<0$, namely, $\phi_J$ would not be ordered by itself. The bare ordering temperature for the nuclear spins is expected to be the smallest compared to the electronic temperature scales: $|T_I| \ll |T_J|,\ T_\mathrm{AF}$. The bilinear hyperfine coupling between $\phi_J$ and $\phi_I$ will be taken as positive, without loss of generality.
The non-linear couplings include $u_\alpha$, the intra-component quartic couplings,
and $\epsilon$ and $\eta$, the inter-component biquadratic couplings.
We will consider all these quartic couplings to be positive, so that the Landau theory for
each component is well-defined and, furthermore, there is a phase competition.

\begin{figure}
	\begin{center}
		\includegraphics[width=0.7\columnwidth]{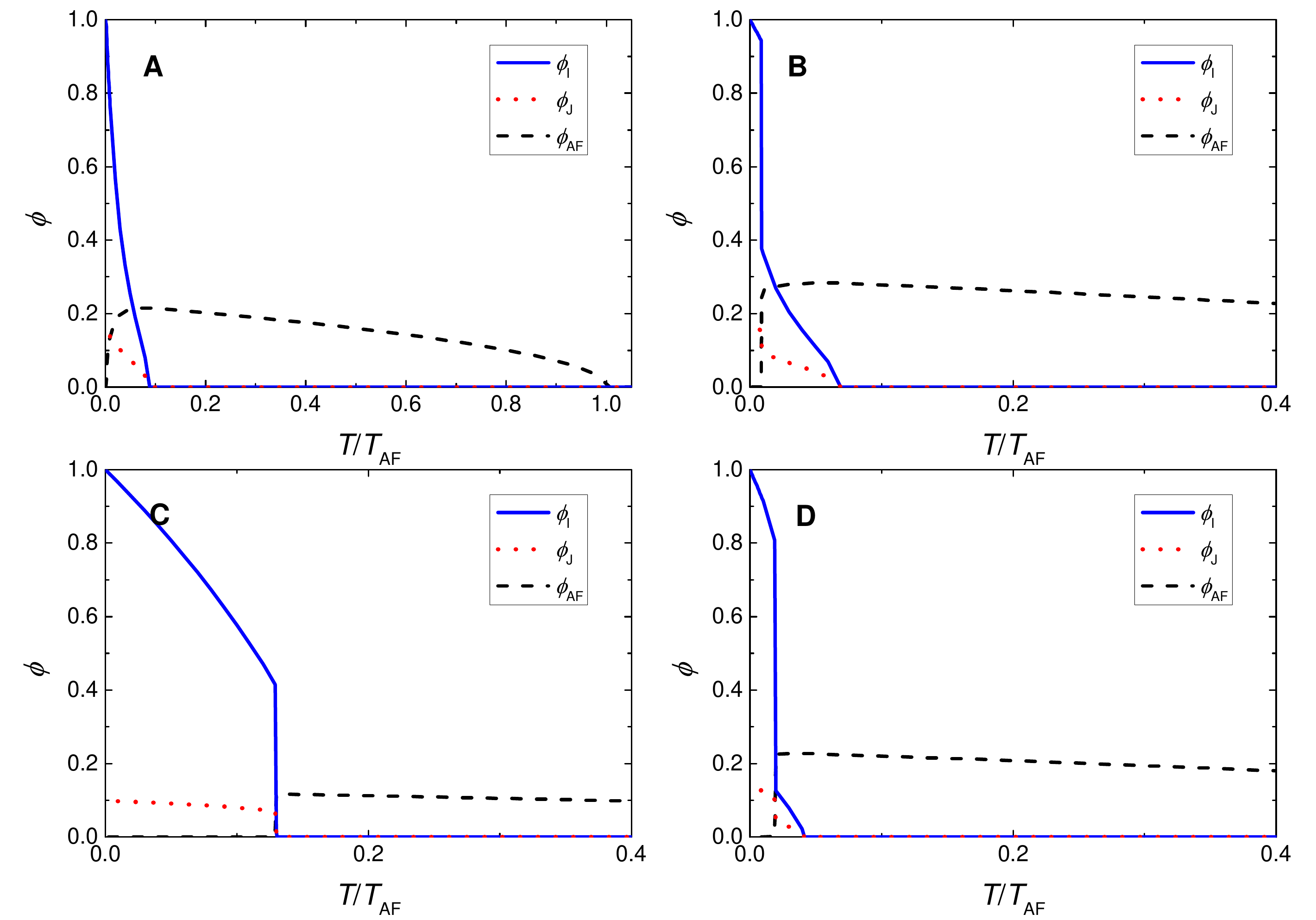}
	\end{center}
	\caption{\label{figS9} Typical temperature evolutions of order parameters in the three-component Landau theory for various model parameters. As shown in panels A-C, three different kinds of evolutions (see text) are found. The most experimentally relevant scenario is shown in panel D, where $T_0\sim T_{hyb}\gg T_I$ (see text for the definition of these temperatures). In each panel, $\phi_\alpha$ ($\alpha=\mathrm{AF},\ J, \ I$) have been rescaled such that $\phi_I(T=0)=1$. The model parameters used for this plot are $T_\mathrm{AF}=1$ (the energy unit), $T_I=-0.001$, $T_J=-0.7$, $u_\mathrm{AF}=2$, $u_J=6.4$. In addition, for panel A, $u_I=0.005$, $\lambda=0.35$, $\eta=0.2$, and $\epsilon=0.1$; for panel B, $u_I=0.005$, $\lambda=0.4$, $\eta=0.3$, and $\epsilon=0.1$; for panel C, $u_I=0.005$, $\lambda=0.3$, $\eta=0.3$, and $\epsilon=0.5$; and for panel D, $u_I=0.002$, $\lambda=0.36$, $\eta=0.2$, and $\epsilon=0.1$.}\label{figST}
\end{figure}

Since the model has two $Z_2$ symmetries, there can be two phase transitions at $T_\mathrm{AF}$ and $T_{hyb}$ corresponding to the ordering of the main electronic spin component $\phi_{\mathrm{AF}}$ and the hybridized electron and nuclear spin, respectively. They can be determined by solving the saddle-point equations:
\begin{eqnarray}
\partial f/\partial \phi_\mathrm{AF} &=& (r_\mathrm{AF}+\eta \phi^2_J + \epsilon \phi^2_I) \phi_\mathrm{AF} + u_\mathrm{AF} \phi^3_\mathrm{AF} = 0\label{Eq:SPE3a}\\
\partial f/\partial \phi_J &=& (r_J+\eta \phi^2_\mathrm{AF}) \phi_J + u_J \phi^3_J - \lambda \phi_I = 0\label{Eq:SPE3b}\\
\partial f/\partial \phi_I &=& (r_I+\epsilon \phi^2_\mathrm{AF}) \phi_I + u_I \phi^3_I - \lambda \phi_J = 0.\label{Eq:SPE3c}
\end{eqnarray}
With the lowering of temperature, the first transition takes place at $T_\mathrm{AF}=70$ mK, and is second-order. At $T<T_\mathrm{AF}$, $\phi_\mathrm{AF}>0$ but $\phi_J=\phi_I=0$, hence $T_\mathrm{AF}$ is not affected by the nuclear spin component $\phi_I$. A hybrid electron and nuclear spin order is stabilized at a lower temperature $T_{hyb}$. This corresponds to $\phi_I>0$ and $\phi_J>0$.
We find that $T_{hyb}\gg T_I$ (see below), and at $T<T_{hyb}$, $\phi_\mathrm{AF}$ decreases with lowering temperature and is substantially suppressed (see Fig.~S\ref{figST}). In this temperature regime, the dominant order parameter can be the hybridized spin order, with its primary component coming from nuclear spins.

At the mean-field level, there could be a third phase transition at $T_0$, below which $\phi_\mathrm{AF}=0$, but the hybridized electron and nuclear spin order is still stabilized. Depending on the model parameters, the transitions at $T_{hyb}$ and $T_0$ can be either first-order or second-order. Here, we will not give an exhaustive discussion on the full phase diagram of this three-component model. We will however show that three different kinds of behavior can be obtained without fine tuning the model parameters in the physical regime. First, as shown in Fig.~S\ref{figST}A, both transitions at $T_{hyb}$ and $T_0$ are second-order. Second, as shown in Fig.~S\ref{figST}B, the transition at $T_{hyb}$ is second-order, but the one at $T_0$ is first-order. Finally, $T_{hyb}=T_0$, and both transitions are first-order, as shown in Fig.~S\ref{figST}C.

$T_{hyb}$ and $T_0$, as well as the order of transition at these temperatures can be determined from a further simplified effective two-component model. To see this, note that Eq.~\eqref{Eq:SPE3c} can be rewritten as $\phi_J = \phi_I (r_I + u_I \phi^2_I + \epsilon \phi^2_\mathrm{AF})/\lambda$. Plug it into the free energy of the three component model, we obtain an effective two-component model for $\phi_\mathrm{AF}$ and $\phi_I$:
\begin{eqnarray}\label{Eq:FE3to2}
f_{\rm{eff}} &=& \frac{r_\mathrm{AF}}{2} \phi^2_\mathrm{AF} + \frac{\tilde{r}_I}{2} \phi^2_I + \frac{u_\mathrm{AF}}{4} \phi^4_\mathrm{AF} + \frac{\tilde{u}_I}{4} \phi^4_I \nonumber\\
&&+\frac{\tilde{\epsilon}}{2} \phi^2_\mathrm{AF} \phi^2_I + O(\phi^6),
\end{eqnarray}
where
\begin{eqnarray}
\tilde{r}_I &=& r_I(\frac{r_J r_I}{\lambda^2}-1)\\
\tilde{u}_I &=& \frac{u_J r^4_I}{\lambda^4}+4\frac{r_J r_I}{\lambda^2} - 3u_I\\
\tilde{\epsilon} &=& \frac{\eta r^2_I}{\lambda^2} + 2\epsilon\frac{r_J r_I}{\lambda^2} - \epsilon.
\end{eqnarray}
Minimizing this effective free energy, which is equivalent to minimize the free energy
given in Eq.~\eqref{FE2}, we find the transition at $T_{hyb}$ is second-order when
$\tilde{\epsilon}^2<u_\mathrm{AF}\tilde{u}_I$ and first-order if $\tilde{\epsilon}^2>u_\mathrm{AF}\tilde{u}_I$. By requiring $\phi_I\to0$ but $\phi_\mathrm{AF}\neq0$ at the minimized free energy, we obtain the equation for $T_{hyb}$. Similarly, we solve for $T_0$ by requiring $\phi_\mathrm{AF}\to0$ but $\phi_I\neq0$ at the minimized free energy. The equations for $T_{hyb}$ and $T_0$ are more involved,  but they can be solved numerically.

In general, we obtain $T_{hyb}\gg T_I$ (see below); $T_0$ is sensitive to the renormalized model parameters $\tilde{r}_I$, $\tilde{u}_I$, $\tilde{\epsilon}$, and can vary in a wide range between $T_I$ and $T_{hyb}$. As shown in Fig.~S\ref{figST}D, without drastic change of the model parameters, we can obtain $T_0\sim T_{hyb}\gg T_I$. This scenario could be the most experimentally relevant once the effects of fluctuations are taken into account. Fluctuations beyond the Landau theory may smear the transition at $T_{hyb}$ into a crossover, and make the true second-order transition to be closer to where there is a rapid increase of the hybrid order (both $ \phi_I $ and $\phi_J$), which is also close to $T_0$. Compared to the experiments, these would respectively correspond to the crossover at $T_B$, below which fluctuations suppressing the primary electronic spin order are observed as an increase of the $4f$-electron spin susceptibility $\chi(T)$, and the second-order transition at $T_A$, at which the nuclear spin order sets in. The additional increase of the measured $M(T)$ reflecting the expected additional drop of $m_\mathrm{AF}$ upon cooling to below \TA\ appears to be almost compensated, even slighty overcompensated, by the ordering of the (1 -- 2\,\%) electronic $\phi_J$ component of the hybrid A-phase. This results in an almost flat, slightly decreasing magnetization below \TA, see Fig.~S3B.

\subsubsection{Hybrid nuclear and electronic spin order below $T_{hyb}$}
If either $\phi_\mathrm{AF}=0$ below $T_{hyb}$ or $\phi_\mathrm{AF}$ varies slowly across $T_{hyb}$, we can describe the transition at $T_{hyb}$ by a simpler Landau theory involving only $\phi_I$ and $\phi_J$. Rewriting Eq.~\eqref{FE2}, the free energy functional is
\begin{eqnarray}
f &=& \frac{r_J^\prime}{2} \phi^2_J + \frac{u_J}{4} \phi^4_J + \frac{r_I^\prime}{2} \phi^2_I + \frac{u_I}{4} \phi^4_I -\lambda \phi_J \phi_I,
\end{eqnarray}\label{FE2c}
where $r_J^\prime=T-T_J^\prime=T-(T_J-\eta\phi_\mathrm{AF}^2)$ and $r_I^\prime=T-T_I^\prime=T-(T_I-\epsilon\phi_\mathrm{AF}^2)$. Here we have fixed $\phi_\mathrm{AF}$ to be a constant.

By solving the saddle point equations, we find that the transition to a hybrid electronic and nuclear spin order of $\phi_J$ and $\phi_I$ is at
\begin{equation}
T_{hyb} = \frac{(T_J^\prime+T_I^\prime)+\sqrt{(T_J^\prime-T_I^\prime)^2+4\lambda^2}}{2}.
\end{equation}
Especially,
in the case we consider, with the $\phi_J$ not being ordered on its own and therefore
the
bilinear hyperfine coupling satisfing
$0<\lambda\ll -T_J^\prime$,
and with $|T_I^{\prime}| \ll -T_J^{\prime}$, we have
\begin{equation}\label{Eq:Thyb}
T_{hyb} \approx T_I^\prime + \frac{\lambda^2}{|T_J^\prime|}. 
\end{equation}

It is interesting to note that in this limit, the enhancement of the nuclear spin ordering temperature $T_{hyb}-T_I^\prime=\lambda^2/|T_J^\prime|$, is independent of $T_I^\prime$. This factor can be physically understood as an effective exchange interaction between the nuclear spins that are mediated by the gapped electronic spin excitations. It is also remarkable that for $\lambda^2> | T_I^\prime T_J^\prime|$, $T_{hyb}>0$ even if $T_I^\prime<0$. In other words, the hyperfine coupling $\lambda$ can induce a hybrid nuclear and electronic spin order even if the two components are not ordered by themselves. At $T<T_{hyb}$, the primary order parameter of this hybrid order can be the nuclear spin order $\phi_I$, as shown in Fig.~S\ref{figST}.

We now estimate $T_{hyb}$ for \YRS. First, the relevant nuclear spin degree of freedom is the Yb nuclear spin because of its strong hyperfine coupling. For Yb, the hyperfine coupling constant $A_{\mathrm{hf}}\sim10^2$ T/$\mu_{\mathrm{B}}$, which gives $\lambda=A_{\mathrm{hf}}g_{el}g_{I}\sim25$ mK. Second, the coupling between the Yb nuclear spins without coupling to the Yb 4$f$ electronic spins is expected to be very small. For simplicity, we will assume $T_I^\prime=0$ for this system. Third, $T_J^{\prime}$ is expected to take a value typically of the spins of the $4f$ electrons. Without loss of generality, we estimate $T_J^{\prime}$ by $g^2_{el}/\chi(\boldsymbol{Q}_1)$, where $\chi(\boldsymbol{Q}_1)$ is the electronic spin susceptibility at wavevector $\boldsymbol{Q}_1$. Assuming that $\chi(\boldsymbol{Q}_1)$ is on the same order as $\chi(\boldsymbol{Q}=0)\sim 1$ $\mu_{\mathrm{B}}$/T, we obtain $T_J^\prime\sim 600$ mK. With these values in place, from Eq.~\eqref{Eq:Thyb}, we obtain $T_{hyb}\sim 1$ mK. This temperature scale is consistent with the experimentally observed transition temperature $T_A$. Moreover, the effective $g$-factor
\begin{eqnarray}
g_{eff} &=& \frac{g_I \phi_I + g_{el} \phi_J}{\phi_I+\phi_J} \nonumber\\
&\approx& g_{el} \frac{\phi_J}{\phi_I+\phi_J}.
\end{eqnarray}
For $T\lesssim T_{hyb}$, $\phi_J/\phi_I\sim \lambda/|T_J^\prime|\sim 0.04$. This gives $g_{eff}/g_{el}\sim 0.04$,
which is also consistent with the experimental observation in terms of order of magnitude.

We close with a caution that the two Yb isotopes containing non-zero nuclear moments have a total abundance of about 30.4\,\%. This is smaller than the percolation threshold ($p_c$) of either a three-dimensional simple cubic lattice ($p_c\approx0.3116$) or a two-dimensional square lattice ($p_c\approx0.592$). So, the nature of $m_I$ should reflect the random distribution of the Yb nuclear spins on the Yb chemical lattice. The specification of this order parameter, however, is beyond the scope of our consideration.
\subsection{Possible coupling of electronic and nuclear spin orders in other heavy-electron metals}
In the weak heavy-electron antiferromagnet UPt$_3$ ($T_{\textrm{N}} \approx 5$\,K, Ref.~\onlinecite{Aeppli}) a phase transition with a large specific-heat anomaly has been discovered at 18\,mK (Ref.~\onlinecite{Schuberth92}) which may indeed indicate the formation of such a hybrid order. Its DC-magnetization vs. $T$ curve has the same shape as is observed for \YRS\ here. UPt$_3$ is a heavy Fermi liquid, in which superconductivity forms at \TC\,$\approx 0.5 K$, i.e., way above this 18\,mK transition and obviously decoupled from it. In the heavy Fermi liquid CeCu$_6$, a phase transition occurs at about 3\,mK (Refs.~\onlinecite{Schuberth95,Pollack}) which was shown to be of AF nature~\cite{Tsujii}. In addition, Schuberth \textit{et al.}~\cite{Schuberth95} found a second anomaly, but no superconductivity, at $T \approx$ 0.6\,mK. It is possible that these anomalies in CeCu$_6$ are related to the formation of electronic antiferromagnetism at 3\,mK, followed by a transition into a hybrid electronic-nuclear spin order at $T \approx 0.6$\,mK. However, since both systems have a weaker hyperfine coupling, the hybrid order probably has a stronger electronic and weaker nuclear character than in \YRS.
\subsection{\YRS, the second Yb-based heavy-electron superconductor: Size of \TC\ and implications for other classes of unconventional superconductors}
\YRS\ is only the second Yb-based heavy-electron superconductor, following $\beta$-YbAlB$_{4}$ with \TC\ = 80\,mK (Ref.~\onlinecite{Nakatsuji}). The latter \TC\ is much smaller than the typical \TC\ of Ce-based heavy-electron superconductors ($\lesssim 2.5$\,K). Such a ratio is also found for the magnetic ordering temperatures, e.g., \TN\ = 70\,mK for \YRS\ compared to \TN\ $\lesssim 4$\,K in CeRhIn$_{5}$; this is commonly attributed to the so-called \textit{lanthanide contraction} of the heavy compared to the light rare earths.

The value of \TC\ = 2\,mK found for \YRS\ is smaller than the highest \TC\ yet observed for an Yb-based superconductor by about a factor of 40. On the other hand, the same spread of \TC s is known for the heavy-electron superconductors based upon light lanthanides, ranging between \TC\ = 2.5\,K for CeAu$_{2}$Si$_{2}$ (Ref.~\onlinecite{Ren}) and \TC\ = 50\,mK for PrIr$_{2}$Zn$_{20}$ (Ref.~\onlinecite{Onimaru}). This means that the \TC\ value of \YRS\ is in the range that can be expected for a heavy-electron metal containing a heavy rare-earth constituent.

Also, the \TC\ of \YRS\ is way above of what is expected from the Kohn-Luttinger hypothesis, according to which each metal may become superconducting at sufficiently low temperature~\cite{Kohn-Luttinger}. As the effective Fermi temperature \TF\ $\simeq T_{\rm K} \simeq 25$\,K, for \YRS, \TC/\TF\ is of the order of 10$^{-4}$, which is unlikely high in the Kohn-Luttinger sense. Most importantly, the Kohn-Luttinger hypothesis does not involve the action of any nuclear spins which however, as it is shown here, are crucial for the development of superconductivity in \YRS: The experimental data presented in this paper, reveal that superconductivity occurs just below the nuclear spin order, and our three-component Ginzburg-Landau theory (see Sec.~\ref{theory}) explains that it is this nuclear spin order which essentially weakens the pair-breaking primary electronic magnetic order (\TN\ = 70\,mK) and pushes the material into the close vicinity of, if not at, its QCP.

Apparently the superconductivity is due to the $B = 0$ quantum criticality induced by the nuclear spin order. Previous thermodynamic measurements have shown that the quantum critical behavior near $B_{\textrm{c}}$ is very similar to that occurring at $B = 0$. For instance, prior to the AF order, i.e.,  at $T >$ \TN\ $\simeq 70$\,mK, $C_{\textrm{el}}(T)/T$ is essentially the same at $B = 0$ and $B = B_{\textrm{c}}$ (Ref.~\onlinecite{Oeschler}). Given the extensive experimental evidence of the Kondo destroying nature of the field-driven QCP in \YRS, we conclude that the superconductivity observed here is likely driven by fermionic critical fluctuations associated with the Kondo destroying QCP (Ref.~\onlinecite{Pixley}). We note that alternative theoretical proposals have been made for the quantum criticality in \YRS\ (Refs.~\onlinecite{Watanabe,Woelfle}). These proposals, however, do not contain the physics associated with the observed $T^{*}(B)$ line for the rapid crossover of the 
Fermi surface.

Superconductivity in heavy-electron metals is often discussed in terms of an effective electron-electron attractive interaction provided by nearly quantum critical fluctuations associated with a spin-density wave (SDW) QCP~\cite{Mathur,Monthoux}. This was recently exemplified, via inelastic neutron scattering, for CeCu$_2$Si$_2$ (Refs.~\onlinecite{Arndt,Stockert}). On the other hand, in the special case of CeRhIn$_5$, superconductivity appears to form~\cite{Shishido,Park,Knebel08} in the vicinity of a Kondo-breakdown QCP~\cite{Si01,Coleman,Senthil}. This is in contrast to the behavior of CeCu$_{6-x}$Au$_x$, the prototype heavy-electron metal which exhibits such a Kondo breakdown QCP~\cite{Loehneysen,Schroeder} but shows no superconductivity down to $T \approx 20$\,mK (Ref.~\onlinecite{Loehneysen95}). In this case, it is natural to assume that unconventional superconductivity is, at least above 20 mK, suppressed by the alloying-induced disorder. By contrast, in high-quality single crystals of the antiferromagnet \YRS, another well-established heavy-electron metal with a Kondo breakdown QCP~\cite{Paschen,Friedemann}, our work shows that superconductivity does develop at \TC\ $= 2$\,mK. Here, the primary electronic order which appears to be detrimental to superconductivity is sufficiently weakened by the ordering of the nuclear spins; this in turn pushes the system close to the underlying QCP. The concomitant quantum critical fluctuations, rather than the magnon fluctuations as in the case of UPd$_{2}$Al$_{3}$ (Ref.~\onlinecite{Sato}), are therefore the driving force for superconductivity. This heavy-electron superconductivity may be called "high \TC", in the sense that it is limited by an exceedingly high ordering temperature of nuclear spins. Moreover, the emergence of superconductivity in \YRS\ provides evidence for the notion that has been implicated by de-Haas-van-Alphen studies on CeRhIn$_5$ in high pulsed magnetic fields~\cite{Yuan}; namely, superconductivity is robust in the vicinity of such a Kondo-breakdown QCP, which may be considered a zero-temperature $4f$-orbital selective Mott transition. The likely Cooper-pair formation driven by fermionic excitations in \YRS\ has also implications for the unconventional superconductivity in the doped Mott insulators of the cuprates~\cite{cuprates}, organic charge-transfer salts~\cite{Kanoda} and, perhaps, some of the Fe-based superconductors~\cite{Dai} where superconductivity has, up to now, been often ascribed to Cooper pairing mediated by bosonic modes, i.e., AF spin fluctuations~\cite{Scalapino}. Finally, our conclusion that quantum criticality is a robust mechanism for superconductivity pertains to wider settings such as finite-density quark matter~\cite{Alford}.
%
%
\bibliography{yrs_schuberth}

\begin{thebibliography}{10}

\bibitem{cuprates}
P.~A. Lee, N.~Nagaosa, X.-G. Wen, {\it Rev. Mod. Phys.\/} {\bf 78}, 17 (2006).

\bibitem{Loehneysen07}
H.~v. L\"ohneysen, A.~Rosch, M.~Vojta, P.~W\"olfle, {\it Rev. Mod. Phys.\/}
  {\bf 79}, 1015 (2007).

\bibitem{Gegenwart}
P.~Gegenwart, Q.~Si, F.~Steglich, {\it Nature Phys.\/} {\bf 4}, 186 (2008).

\bibitem{Mathur}
N.~D. Mathur, {\it et~al.\/}, {\it Nature\/} {\bf 394}, 39 (1998).

\bibitem{Custers}
J.~Custers, {\it et~al.\/}, {\it Nature\/} {\bf 424}, 524 (2003).

\bibitem{Paschen}
S.~Paschen, {\it et~al.\/}, {\it Nature\/} {\bf 432}, 881 (2004).

\bibitem{Friedemann}
S.~Friedemann, {\it et~al.\/}, {\it Proc. Natl. Acad. Sci. USA\/} {\bf 107},
  14547 (2010).

\bibitem{Ronnow}
H.~M. R{\o}nnow, {\it et~al.\/}, {\it Science\/} {\bf 308}, 389 (2005).

\bibitem{Andres}
K.~Andres, E.~Bucher, P.~H. Schmidt, J.~P. Maita, S.~Darack, {\it Phys. Rev.
  B\/} {\bf 11}, 4364 (1975).

\bibitem{Steinke}
L.~Steinke, {\it et~al.\/}, {\it Phys. Rev. Lett.\/} {\bf 111}, 077202 (2013).

\bibitem{Schuberth13}
E.~Schuberth, M.~Tippmann, C.~Krellner, F.~Steglich, {\it physica status solidi
  (b)\/} {\bf 250}, 482 (2013).

\bibitem{Steppke}
A.~Steppke, {\it et~al.\/}, {\it physica status solidi (b)\/} {\bf 247}, 737
  (2010).

\bibitem{Sichelschmidt}
J.~Sichelschmidt, V.~A. Ivanshin, J.~Ferstl, C.~Geibel, F.~Steglich, {\it Phys.
  Rev. Lett.\/} {\bf 91}, 156401 (2003).

\bibitem{Knebel06}
G.~Knebel, {\it et~al.\/}, {\it J. Phys. Soc. Jpn.\/} {\bf 75}, 114709 (2006).

\bibitem{Monthoux}
P.~Monthoux, D.~Pines, G.~G. Lonzarich, {\it Nature\/} {\bf 450}, 1177 (2007).

\bibitem{Arndt}
J.~Arndt, {\it et~al.\/}, {\it Phys. Rev. Lett.\/} {\bf 106}, 246401 (2011).

\bibitem{Stockert}
O.~Stockert, {\it et~al.\/}, {\it Nature Phys.\/} {\bf 7}, 119 (2011).

\bibitem{Shishido}
H.~Shishido, R.~Settai, H.~Harima, Y.~\={O}nuki, {\it J. Phys. Soc. Jpn.\/}
  {\bf 74}, 1103 (2005).

\bibitem{Park}
T.~Park, {\it et~al.\/}, {\it Nature\/} {\bf 440}, 65 (2006).

\bibitem{Knebel08}
G.~Knebel, D.~Aoki, J.-P. Brison, J.~Flouquet, {\it J. Phys. Soc. Jpn.\/} {\bf
  77}, 114704 (2008).

\bibitem{Si01}
Q.~Si, S.~Rabello, K.~Ingersent, J.-L. Smith, {\it Nature\/} {\bf 413}, 804
  (2001).

\bibitem{Coleman}
P.~Coleman, C.~P{\'e}pin, Q.~Si, R.~Ramazashvili, {\it J. Phys.: Condens.
  Matter\/} {\bf 13}, R723 (2001).

\bibitem{Senthil}
T.~Senthil, M.~Vojta, S.~Sachdev, {\it Phys. Rev. B\/} {\bf 69}, 03511 (2004).

\bibitem{Nakatsuji}
S.~Nakatsuji, {\it et~al.\/}, {\it Nature Phys.\/} {\bf 4}, 603 (2008).

\bibitem{Loehneysen}
H.~v. L\"ohneysen, {\it et~al.\/}, {\it Phys. Rev. Lett.\/} {\bf 72}, 3262
  (1994).

\bibitem{Schroeder}
A.~Schr\"oder, {\it et~al.\/}, {\it Nature\/} {\bf 407}, 351 (2000).

\bibitem{Loehneysen95}
H.~v. L\"ohneysen, {\it Physica B\/} {\bf 206--207}, 101 (1995).

\bibitem{Sato}
N.~K. Sato, {\it et~al.\/}, {\it Nature\/} {\bf 410}, 340 (2001).

\bibitem{Yuan}
L.~Jiao, {\it et~al.\/}, {\it Proc. Natl. Acad. Sci. USA\/} {\bf 112}, 673
  (2015).

\bibitem{Kanoda}
K.~Kanoda, {\it Mott transition and superconductivity in quasi-2D organic
  conductors\textnormal{, in A. Lebed (ed.) {"}The Physics of Organic
  Superconductors and Conductors{"}}\/} (Springer-Verlag, Berlin Heidelberg,
  2008).

\bibitem{Alford}
M.~Alford, A.~Schmitt, K.~Rajagopal, T.~Sch\"afer, {\it Rev. Mod. Phys.\/} {\bf
  80}, 1455 (2008).

\bibitem{SOM}
Materials and methods are available as supplementary materials on science
  online.

\bibitem{Steglich}
F.~Steglich, {\it et~al.\/}, {\it Phys. Rev. Lett.\/} {\bf 43}, 1892 (1979).

\bibitem{Si2010}
Q.~Si, F.~Steglich, {\it Science\/} {\bf 329}, 1161 (2010).

\bibitem{Gegenwart07}
P.~Gegenwart, {\it et~al.\/}, {\it Science\/} {\bf 315}, 969 (2007).

\bibitem{Ishida03}
K.~Ishida, {\it et~al.\/}, {\it Phys. Rev. B\/} {\bf 68}, 184401 (2003).

\bibitem{Rauchschwalbe}
U.~Rauchschwalbe, {\it et~al.\/}, {\it Phys. Rev. Lett.\/} {\bf 49}, 1448
  (1982).

\bibitem{Ott}
H.~R. Ott, H.~Rudigier, Z.~Fisk, J.~L. Smith, {\it Phys. Rev. Lett.\/} {\bf
  50}, 1595 (1983).

\bibitem{Langhammer}
C.~Langhammer, {\it et~al.\/}, {\it J. Magn. Magn. Mat.\/} {\bf 177--181, Part
  1}, 443 (1998).

\bibitem{Assmus}
W.~Assmuss, {\it et~al.\/}, {\it Phys. Rev. Lett.\/} {\bf 52}, 469 (1984).

\bibitem{Rauchschwalbe87}
U.~Rauchschwalbe, {\it Physica\/} {\bf 147B}, 1 (1987).

\bibitem{Schuberth84}
E.~Schuberth, {\it Rev. Sci. Instr.\/} {\bf 55}, 1486 (1984).

\bibitem{Krellner}
C.~Krellner, S.~Taube, T.~Westerkamp, Z.~Hossain, C.~Geibel, {\it Phil. Mag.\/}
  {\bf 92}, 2508 (2012).

\bibitem{Lausberg}
S.~Lausberg, {\it et~al.\/}, {\it Phys. Rev. Lett.\/} {\bf 110}, 256402 (2013).

\bibitem{Westerkamp}
T.~Westerkamp, P.~Gegenwart, C.~Krellner, C.~Geibel, F.~Steglich, {\it Physica
  B\/} {\bf 403}, 1236 (2008).

\bibitem{Feyerherm}
R.~Feyerherm, {\it et~al.\/}, {\it Phys. Rev. B\/} {\bf 56}, 699 (1997).

\bibitem{Brando}
M.~Brando, {\it et~al.\/}, {\it physica status solidi (b)\/} {\bf 250}, 485
  (2013).

\bibitem{Kalvius}
G.~M. Kalvius, G.~K. Shenoy, B.~D. Dunlap, {\it Coll. Int. Cent. Nat. Rech.
  Sci.\/} {\bf 2}, 477 (1970).

\bibitem{Shimizu}
T.~Shimizu, M.~Takigawa, H.~Yasuoka, J.~Wernick, {\it J. Magn. Magn. Mat.\/}
  {\bf 52}, 187 (1985).

\bibitem{Seitchik}
J.~A. Seitchik, V.~Jaccarino, J.~H. Wernick, {\it Phys. Rev.\/} {\bf 138}, A148
  (1965).

\bibitem{Ishida}
K.~Ishida, {\it et~al.\/}, {\it Phys. Rev. Lett.\/} {\bf 89}, 107202 (2002).

\bibitem{Aeppli}
G.~Aeppli, {\it et~al.\/}, {\it Phys. Rev. Lett.\/} {\bf 63}, 676 (1989).

\bibitem{Schuberth92}
E.~A. Schuberth, B.~Strickler, K.~Andres, {\it Phys. Rev. Lett.\/} {\bf 68},
  117 (1992).

\bibitem{Schuberth95}
E.~A. Schuberth, J.~Schupp, R.~Freese, K.~Andres, {\it Phys. Rev. B\/} {\bf
  51}, 12892 (1995).

\bibitem{Pollack}
L.~Pollack, {\it et~al.\/}, {\it Phys. Rev. B\/} {\bf 52}, R15707 (1995).

\bibitem{Tsujii}
H.~Tsujii, {\it et~al.\/}, {\it Phys. Rev. Lett.\/} {\bf 84}, 5407 (2000).

\bibitem{Ren}
Z.~Ren, {\it et~al.\/}, {\it Phys. Rev. X\/} {\bf 4}, 031055 (2014).

\bibitem{Onimaru}
T.~Onimaru, {\it et~al.\/}, {\it J. Phys. Soc. Jpn.\/} {\bf 79}, 033704 (2010).

\bibitem{Kohn-Luttinger}
W.~Kohn, J.~M. Luttinger, {\it Phys. Rev. Lett.\/} {\bf 15}, 524 (1965).

\bibitem{Oeschler}
N.~Oeschler, {\it et~al.\/}, {\it Physica B\/} {\bf 403}, 1254 (2008).

\bibitem{Pixley}
J.~H. Pixley, L.~Deng, K.~Ingersent, Q.~Si, {\it Phys. Rev. B\/} {\bf 91},
  201109 (2015).

\bibitem{Watanabe}
S.~Watanabe, K.~Miyake, {\it J. Phys. Soc. Jpn.\/} {\bf 82}, 083704 (2013).

\bibitem{Woelfle}
P.~W\"olfle, E.~Abrahams, {\it Phys. Rev. B\/} {\bf 84}, 041101 (2011).

\bibitem{Dai}
J.~Dai, Q.~Si, J.-X. Zhu, E.~Abrahams, {\it Proc. Natl. Acad. Sci. USA\/} {\bf
  106}, 4118 (2009).

\bibitem{Scalapino}
D.~J. Scalapino, {\it Rev. Mod. Phys.\/} {\bf 84}, 1383 (2012).

\end{thebibliography}
\bibliographystyle{Science}
\end{document}